\documentclass[letterletter,10pt,journal, twoside]{IEEEtran}
\input epsf
\usepackage[square,numbers,sort]{natbib}
\usepackage{xcolor,soul,framed} %,caption
\colorlet{shadecolor}{yellow}
\usepackage{graphicx}
\usepackage{amsmath}
\newtheorem{theorem}{\textbf{Theorem}}
\newtheorem{lemma}{\textbf{Lemma}}
\newtheorem{example}{\textbf{Example}}
\newtheorem{corollary}{\textbf{{Corollary}}}
\newtheorem{remark}{\textbf{Remark}}
\newtheorem{definition}{\textbf{Definition}}

\newtheorem{proposition}{\textbf{Proposition}}

\usepackage{url}
	\newenvironment{proof}{{{\bf Proof:}}}{\hfill $\square$\par}
	\usepackage{multirow} %multirow for format of table
	\usepackage{xcolor}
	\usepackage{multirow}
	\usepackage{setspace}
	\usepackage{url}
	\usepackage{subfigure}
	\usepackage{fancyhdr}
	\usepackage{amsmath}
	\usepackage{multirow}
	\usepackage{amssymb }
	\usepackage{color}
	\usepackage{graphics} % for pdf,  bitmapped graphics files
	\usepackage{graphicx}
	\usepackage{algorithm} %format of the algorithm
	\usepackage{algorithmic} %format of the algorithm
	\usepackage{subeqnarray}
	\usepackage{cases}
	\usepackage{setspace}

	\renewcommand{\baselinestretch}{1.1} \normalsize
\begin{document}
			
			%\title{Submission Format for IMS2014 (Title in 24-point Times font)}
			% If the \LARGE is deleted, the title font defaults to  24-point.
			% Actually,
			% the \LARGE sets the title at 17 pt, which is close enough to 18-point.
			%+++++++++++++++++++++++++++++++++++++++++++ LP-based Method and
			% \title{{\LARGE Total Unimodularity and Polynomial Solvability of Several Cost-Sparsity Induced Input Selection Problems for Structural Controllability: an LP-based Method}}
			%	\title{{\Large On Constrained Input Selections for Structured Systems: Polynomially Solvable Cases}}
			%	\title{\LARGE On Polynomially Solvable Cases of Constrained Input Selections \\~ for Structured Network Systems: an LP-based Method}
			%% On Polynomially Solvable Cases of Constrained Input Selections for Time-invariant and Swithced Structured Systems
			%% On Polynomially Solvable Cases of Constrained Input Selections for Structured Networks with Fixed and Swithed Topologies
			%		\title{\LARGE Comments on structural controllability of switched linear systems}
			%	On the proof for structural controllability of switched linear systems
			
			%	\title{\LARGE Structural controllability of switched linear systems revisited}
	%		\title{\LARGE Structural Controllability of Switched Continuous and Discrete Time\\~ Linear Systems}
			%	\title{\LARGE Structural Controllability of {Switched Continuous-Time Systems Revisited}}
				\title{\LARGE {Generalized Cactus} and Structural Controllability of Switched Linear {Continuous-Time} Systems}
			
			%\author{\authorblockN{J. Clerk Maxwell}
				%\authorblockA{School of Electrical and\\Computer Engineering\\
					%Somewhere Institute of Technology\\
					%City, State 54321--0000\\
					%Email: maxwell@curl.edu}
				%\and
				%\authorblockN{Michael Faraday}
				%\authorblockA{(List authors on this line using 12 point Times font\\ - use a second line if necessary)\\
					%Microwave Research\\
					%City, State/Region, Mail/Zip Code, Country\\
				%\and
				%\authorblockN{Andr\'e M. Amp\`ere \\ }
				%\authorblockA{Starfleet Academy\\
					%San Francisco, CA 96678-2391\\
			
			\author{Yuan Zhang, Yuanqing Xia, and {Aming Li}% <-this % stops a space %, Yuanqing Xia, and Yufeng Zhan
				
				%  \thanks{This work was supported in part by the China Postdoctoral Inno-
					%vative Talent Support Program under Grant BX20200055, the China
					%Postdoctoral Science Foundation under Grant 2020M680016, and the
					%National Natural Science Foundation of China under Grant 62003042.
					%}
				\thanks{This work was supported in part by the
					National Natural Science Foundation of China under Grant 62373059. Y. Zhang and Y. Xia are with School of Automation, Beijing Institute of Technology, Beijing, China. {A. Li is with Center for Systems and Control, College of Engineering, Peking University, Beijing, China.} Email: zhangyuan14@bit.edu.cn, xia\_yuanqing@bit.edu.cn, amingli@pku.edu.cn. }% <-this % stops a space %Email: {zhangyuan14@bit.edu.cn}.
			}
			\maketitle %Given is an autonomous system and a constrained input configuration where whether an input can directly actuate a state variable, as well as the corresponding
			%cost, is prescribed, the problems include, on the prescribed input configurations
			{\small{		{
						\begin{abstract} This paper explores the structural controllability of switched linear continuous-time systems. It first identifies a gap in the proof for a pivotal criterion for the structural controllability of switched linear systems in the literature. To address this void, we develop novel graph-theoretic concepts, such as multi-layer dynamic graphs, generalized stems/buds, and generalized cacti, and based on them, provide a comprehensive proof for this criterion. Our approach also induces a new, generalized cactus based graph-theoretic criterion for structural controllability. This not only extends Lin's cactus-based graph-theoretic condition to switched systems for the first time, but also provides a lower bound for the generic dimension of controllable subspaces. %Finally, we present extensions to reversible switched discrete-time systems, which lead to not only a simplified necessary and sufficient condition for structural controllability, but also the determination of the generic dimension of controllable subspaces.

						\end{abstract}
						\IEEEoverridecommandlockouts
						
						% for peerreview papers, inserts a page break and creates the second title.
						% Will be ignored for other modes.
						\IEEEpeerreviewmaketitle

						\section{Introduction} \label{intro-sec}	
						Switched systems represent a class of hybrid systems where multiple subsystems are regulated by switching laws. This dynamic switching among subsystems can enrich the control strategies, often resulting in superior control performance compared to non-switched systems \cite{liberzon1999basic}. Due to its significance in both practical applications and theoretical exploration, the study of switched systems has garnered substantial attention \cite{conner1987structure,Z.S2002Controllability,sun2005analysis}. %For instance, scenarios arise where stabilization through a constant feedback controller remains unattainable, yet becomes feasible through transitions between distinct constant feedback controllers \cite{narendra1997adaptive}.% For instance, scenarios arise where stabilization through a constant feedback controller remains unattainable, yet becomes feasible through transitions between distinct constant feedback controllers \cite{narendra1997adaptive}.
						
						Controllability and observability are two fundamental concepts that are prerequisite for the design of switched systems. Extensive investigations have been dedicated to these concepts \cite{conner1987structure,ge2001reachability,Z.S2002Controllability,xie2003controllability,ji2009new}. Notably, it has been revealed that the reachable and controllable sets of switched continuous-time systems exist as subspaces within the total space, with complete characterizations established in \cite{Z.S2002Controllability}. However, for discrete-time systems, the reachable and controllable sets do not necessarily manifest as subspaces \cite{stanford1980controllability}. Geometric characterizations, ensuring these sets to span the entire space, were introduced in \cite{ge2001reachability}. Simplified criteria for reversible systems can be found in \cite{ji2009new}.
						
						It is important to note that the aforementioned outcomes rely on algebraic computations centered around precise values of system matrices. In practical scenarios, these exact system parameters might be elusive due to modeling inaccuracies or parameter uncertainties \cite{Y.Y.2011Controllability}. When only the zero-nonzero patterns of system matrices are accessible, Liu et al. \cite{LiuStructural} introduced the concept of {\emph{structural controllability}} for {\emph{switched systems}}, aligning with the structural controllability concept for linear time-invariant (LTI) systems initiated by \cite{C.T.1974Structural}. In \cite{LiuStructural}, several equivalent criteria, grounded in colored union graphs, were proposed to assess structural controllability. Notably, one criterion involves an input accessibility condition and a generic rank condition, distinguished by its simplicity and elegance. This criterion extends naturally from the structural controllability criterion for LTI systems \cite{generic}. Exploiting this resemblance, elegant outcomes regarding optimal input selections for LTI systems \cite{Ramos2022AnOO} have been extended to switched systems \cite{pequito2017structural,zhang2022constrained}. Liu et al.'s work has also stimulated research exploration into (strong) structural controllability for other classes of time-varying systems, such as linear parameter varying systems \cite{mousavi2020strong},  temporal networks \cite{posfai2014structural,zhang2023reachability}.
						
						Nevertheless, in this paper, we identify a gap in the proof of the criterion's sufficiency. Regrettably, this gap seems unaddressed if we follow the original research thread in \cite{LiuStructural} (refer to Section \ref{sec-II}). Nonetheless, we establish the correctness of the said criterion by providing a rigorous and comprehensive proof for it. Our proof relies on novel graph-theoretic concepts, including multi-layer dynamic graphs, generalized stems, generalized buds, and generalized cactus configurations. Our approach also births a new criterion for structural controllability based on generalized stem-bud structures (namely, generalized cacti). Notably, this extends Lin's cactus-based graph-theoretic condition for structural controllability \cite{C.T.1974Structural} to switched systems for the first time. This criterion also induces a lower bound for the generic dimension of controllable subspaces. %Lastly, we extend these results to reversible switched discrete-time systems. This not only yields simplified necessary and sufficient conditions for structural controllability but also enables us to determine the generic dimension of controllable subspaces.
						
						The rest are organized as follows. Section \ref{sec-II} provides some basic preliminaries and the motivations. Section \ref{sec-III} presents a new generalized cactus configuration based criterion for structural controllability and establishes the correctness of the existing one.  The last section concludes this paper. % Extensions to reversible switched discrete-time systems are given in Section \ref{sec-IV}.
%of this paper by identifying a gap in the existing literature
						
						{{\bf Notations:}  Let ${\bf V}$ be a linear space and ${\bf V}_1$ and ${\bf V}_2$ be subspaces of ${\bf V}$. The sum of ${\bf V_1}$ and ${\bf V}_2$ is defined as ${\bf V}_1+{\bf V}_2=\{v_1+v_2:v_1\in {\bf V}_1,v_2\in {\bf V}_2\}$. Similarly, $\sum\nolimits_{i=1}^k {\bf V}_i$ denotes the sum of $k$ subspaces. Let $\left[X_i|_{i=1}^k\right]$ be a composite matrix whose column blocks are $X_1,...,X_k$, which may be occasionally written as $[X_1,...,X_k]$. The column space spanned by a matrix $M$ is denoted by ${\rm span} M$. For an integer $N\ge 1$, $[N]\doteq \{1,...,N\}$. }
						
						\section{Preliminaries and Motivations} \label{sec-II}
						\subsection{Controllability of switched systems}
						Consider a switched continuous-time linear system whose dynamics is governed by \cite{sun2005analysis}
						\begin{equation} \label{plant-switched}
							\dot x(t)= A_{\sigma(t)} x(t) + B_{\sigma(t)} u_{\sigma(t)}(t),
						\end{equation}
						where $x(t)\in {\mathbb R}^n$ is the state, $\sigma(t): [0,\infty) \to \{1,...,N\}$ is the switching signal that can be designed, $u_{\sigma(t)}(t)\in {\mathbb R}^{m_{\sigma(t)}}$ is the piecewise continuous input, $m_i\in {\mathbb N}$, $i=1,...,N$.
						$( A_i, B_i)$ is called a subsystem of system (\ref{plant-switched}), $i=1,...,N$. $\sigma(t)=i$ implies the subsystem $(A_i, B_i)$ is activated as the system realization at time instant $t$. We may use the matrix set $(A_i,B_i)|_{i=1}^N$ to denote the switched system (\ref{plant-switched}).
						
						\begin{definition}[\cite{Z.S2002Controllability}] \label{def-controllability}
							A state $x\in {\mathbb R}^n$ is said to be controllable, if there exists a finite $t_f$, a switching signal $\sigma(t): [0, t_f)\to \{1,...,N\}$ and an input $u(t) \in {\mathbb R}^{m_{\sigma(t)}}$, $t\in [0, t_f)$, such that $x(0)=x$ and $x(t_f)=0$. The controllable set (controllable subspace) of system (\ref{plant-switched}) is the set of controllable states. System (\ref{plant-switched}) is said to be controllable, if its controllable set is ${\mathbb R}^n$.
						\end{definition}

						Note that if we change `$x(0)=x$ and $x(t_f)=0$' to `$x(0)=0$ and $x(t_f)=x$' in Definition \ref{def-controllability}, then the concepts `reachable set' and `reachability' can be defined. For switched continuous-time systems, their reachable set and controllable set always coincide \cite{Z.S2002Controllability,ge2001reachability}. {Moreover, it has been shown in \cite[Def 7]{Z.S2002Controllability} that the controllability in Definition \ref{def-controllability} is equivalent to the ability to steer the state from any initial state to any other state in ${\mathbb R}^n$ in finite time by suitably choosing the control input and switching signal $\sigma(t)$. }   %The controllability of switched systems can be characterized as follows.
						% and reversible switched discrete time systems (see Section \ref{sec-IV})
						
						\begin{lemma}[\cite{Z.S2002Controllability}] \label{switched-criteria}
							The switched system (\ref{plant-switched}) is controllable, if and only if the following controllability matrix ${\cal R}$ has full row rank:
							\begin{equation}\label{control-m}{\cal R}=\left[A_{i_n}^{j_n}A_{i_{n-1}}^{j_{n-1}}\cdots A_{i_1}^{j_1} B_{i_1}|_{i_1,...,i_n=1,...,N}^{j_n,...,j_1=0,...,n-1}\right],\end{equation}
							{where the column block $A_{i_n}^{j_n}\cdots A_{i_1}^{j_1} B_{i_1}$ runs across all combinations of $i_1,...,i_n,j_1,...,j_n$ satisfying $1\le i_1,...,i_n\le N, 0\le j_1,...,j_n\le n-1$}.
							Moreover, the controllable subspace of system (\ref{plant-switched}) is ${\rm span} {\cal R}$, and its dimension equals ${\rm rank} {\cal R}$.
						\end{lemma}
						
				{A structured matrix is a matrix with entries that are either fixed to be zero or can take arbitrary real values (including zero)  independently (i.e., there is no parameter dependence among them; the latter are called nonzero entries) \cite{generic}.} The {\emph{generic rank}} of a structured matrix (or a polynomial of structured matrices), given by ${\rm grank}(\cdot)$, is the maximum rank it can achieve as a function of parameters for its nonzero entries. It turns out that the generic rank is also the rank this matrix can achieve for almost all values of its nonzero entries. {Here, ``almost all'' means all parameter values except for a set of Lebesgue measure zero in the parameter space. }
						When only the zero-nonzero patterns of matrices $(A_i,B_i)|_{i=1}^N$ are available, that is, $(A_i,B_i)|_{i=1}^N$ are structured matrices, system (\ref{plant-switched}) is called a structured system. $(\tilde A_i,\tilde B_i)|_{i=1}^N$ is a realization of $(A_i,B_i)|_{i=1}^N$, if $\tilde A_i$ ($\tilde B_i$) is obtained by assigning some particular values to the nonzero entries of $A_i$ ($\tilde B_i$), $i=1,...,N$.
						
						\begin{definition}[\cite{LiuStructural}] \label{def-structural-controllability}
							A structured system (\ref{plant-switched}) is said to be structurally controllable, if there is a controllable realization $(\tilde A_i,\tilde B_i)|_{i=1}^N$ of $(A_i,B_i)|_{i=1}^N$.
						\end{definition}
						
						{From \cite{LiuStructural}}, if a system is structurally controllable, then almost all its realizations are controllable. This generic property makes this concept appealing, especially for analyzing large-scale network systems with known typologies but unknown interconnection weights~\cite{Y.Y.2011Controllability}.
						
						\subsection{Graph-theoretic preliminaries}
						A directed graph (digraph) is denoted by ${\cal G}=(V,E)$, where $V$ is the vertex set, and $E\subseteq V\times V$ is the edge set. A {\emph{subgraph}} of ${\cal G}$ is a graph ${\cal G}_s=(V_s,E_s)$ such that $V_s\subseteq V$ and $E_s\subseteq E$, and is called a subgraph induced by $V_s$, if $E_s=(V_s\times V_s)\cap E$. We say ${\cal G}_s$ {\emph{covers}} $V_s'\subseteq V$ if $V_s'\subseteq V_s$, and ${\cal G}_s$ {\emph{spans}} ${\cal G}$ if $V_s=V$.   {An edge from $v_{1}$ to $v_{2}$, given by $(v_{1},v_{2})$, is called an {\emph{ingoing edge}} of vertex $v_{2}$, and an {\emph{outgoing edge}} of vertex $v_{1}$.} 	
						A sequence of successive edges {$(v_1,v_2),(v_2,v_3),...,(v_{k-1},v_{k})$} is called a {\emph{walk from vertex $v_1$ to vertex $v_k$}}.  Such a walk $p$ is either denoted by the sequence of edges it contains, i.e., $p=(e_1,e_2,...,e_{k-1})$, where $e_j=(v_{j},v_{j+1})$, or the sequence of vertices it passes, i.e., $p=(v_1,v_2,...,v_k)$. Vertex $v_1$ is called the {\emph{tail}} (initial vertex), denoted as ${\rm tail}(p)$, and vertex $v_k$ is called the {\emph{head}} (terminal vertex), denoted as ${\rm head}(p)$. The {\emph{length}} of a walk $p$, given by $|p|$, is the number of edges it contains (counting repeated edges). A walk {\emph{without repeated vertices}} is called a (simple) {\emph{path}}. A walk from a vertex to itself is called a loop. If the head (or tail) of a loop is the only repeated vertex when traversing along its way, this loop is called a {\emph{cycle}}.

						Two typical graph-theoretic presentations of system (\ref{plant-switched}) are introduced. For the $i$th subsystem, the {\emph{system digraph}} ${\cal G}_i=(X_{i}\cup U_{i},E_{XXi}\cup E_{UXi})$, where the state vertices $X_{i}=\{x^i_1,...,x^i_n\}$, the input vertices $U_{i}=\{u^i_{1},...,u^i_{m_i}\}$, the state edges $E_{XXi}=\{(x^i_k,x^i_j): A_{i,jk}\ne 0\}$, and input edges $E_{UXi}=\{(u^i_{k},x^i_{j}): B_{i,jk}\ne 0\}$. The {\emph{colored union graph}} ${\cal G}_c$ of system (\ref{plant-switched}) is the union of ${\cal G}_1,\cdots, {\cal G}_N$ by using different colors to distinguish state edges from different subsystems. More precisely, ${\cal G}_c$ is a digraph ${\cal G}_c=(X\cup U, E_{XX}\cup E_{UX})$, where $X=\{x_1,...,x_n\}$, $U=U_{1}\cup\cdots\cup U_{N}$, $E_{XX}=\{(x_k,x_j): A_{i,jk}\ne 0, i=1,...,N\}$, and $E_{UX}=\{(u^i_{k},x_j): B_{i,jk}\ne 0,i=1,...,N,k=1,...,m_i\}$. Notice that multiple edges are allowable in $E_{XX}$, and to distinguish them, we assign the {\emph{color index}} $i$ to the edge $(x_k,x_j)$ (resp. $(u_{k},x_j)$) corresponding to $A_{i,jk}\ne 0$ ($B_{i,jk}\ne 0$), $i\in \{1,...,N\}$. An edge $(x_k,x_j)$ with color index $i$ is also denoted by $e_{kj}^i$ (an edge $(u_k^i,x_j)$ with color index $i$ is occasionally denoted by $e_{kj}^i$ if no confusion is made). A {\emph{stem}} is a path from some $u\in U$ to some $x\in X$ in ${\cal G}_c$.
						
						%    A state cactus is a subgraph of ${\cal G}_c$ that consists of a collection of vertex-disjoint stems and cycles, and for each cycle, there is a unique edge from one vertex of the stems to one vertex of the cycle.
						
						\begin{definition}
							A state vertex $x_i\in X$ is said to be input-reachable, if there is a path from an input vertex $u\in U$ to {$x_i$} in ${\cal G}_c$.
						\end{definition}

						\begin{definition}[\cite{LiuStructural}]
							In the colored union graph ${\cal G}_c$, $k$ edges are said to be S-disjoint if their heads are all distinct and if all the edges that have the same tail have different color indices.
						\end{definition}
						
						%     \begin{definition}
							%	A subgraph ${\cal G}_s=(V_s,E_s)$ of the colored union graph ${\cal G}_c$ is said to have a $S$-dilation, if there is a subset $X_s\subseteq $
							%    \end{definition}
						
						The following lemma reveals the relation between the $S$-disjoint edge and ${\rm grank}[A_1,...,A_N, B_1,...,B_N]$.
						
						\begin{lemma}[\cite{LiuStructural}]\label{s-disjoint-edge}
							There are $n$ $S$-disjoint edges in ${\cal G}_c$, if and only if ${\rm grank}[A_1,...,A_N, B_1,...,B_N]=n$.
						\end{lemma}
						
						\subsection{Motivations}
%The recursive formula for verifying the full row rank of ${\mathcal R}$ incurs $O(xx)$ SVD computations. When $n,m$ is large, computationally infeasible and in-stable.  		

{{Inspired by \cite{de2003digital,LiuStructural}, we first provide a practical example of a structured switched linear system.}}

{{\begin{example}Consider a boost converter represented in Fig. \ref{switched-example}(a).
Here, $L$ is an inductance,
$C$ a capacitance, $R$ a load resistance,  $e_S(t)$ the
source voltage, and $i_L(t)$ is the current through the inductance. The switch state $s(t)\in \{0,1\}$. The boost converter can transform the source voltage $e_S(t)$ into a higher voltage $e_C(t)$ over the load $R$ by controlling the switch. The dynamics of the boost converter is  $\dot e_C=-\frac{1}{RC}e_C+(1-s)\frac{1}{C}i_L,\dot i_L=-(1-s)\frac{1}{L}e_C+\frac{1}{L}e_S,$where for notation simplicity, the temporal variable $t$ has been omitted. By taking $[e_C;i_L]$ as the state vector, we obtain a switched linear system model described by (\ref{plant-switched}), with $\sigma(t)=s(t)+1\in \{1,2\}$, and $$A_1=\left[\begin{array}{cc}
	-\frac{1}{RC} & \frac{1}{C}\\
	-\frac{1}{L} & 0
\end{array}\right],A_2=\left[\begin{array}{cc}
	-\frac{1}{RC} & 0\\
	0 & 0
\end{array}\right],B_1=B_2=\left[\begin{array}{c}
0\\
\frac{1}{L}
\end{array}\right].$$
Therefore, it is easy to capture the zero-nonzero patterns of $(A_i,B_i)|_{i=1}^2$, illustrated in Fig. \ref{switched-example}(b). When the exact values of $R,L,C$, as well as the switch signal $s(t)$ that is driven by a pulse-width modulator \cite{de2003digital}, are unknown, this structured system model can tell us whether the structure of the considered electrical network admits controllable realizations (with the flexibility to design $s(t)$). If the answer is yes, its randomized realizations will be controllable with probability one (not counting the parameter dependence among the nonzero entries); otherwise, no controllable realizations exist.
\end{example}}

						Liu et al. \cite{LiuStructural} propose a criterion for the structural controllability of system (\ref{plant-switched}). This criterion says system (\ref{plant-switched}) is structurally controllable, if and only if two conditions hold:
						(i) every state vertex is input-reachale in ${\cal G}_c$, and (ii) ${\rm grank}[A_1,...,A_N, B_1,...,B_N]=n$ (see \citep[Theo 9]{LiuStructural}).
							The necessity of conditions (i) and (ii) is relatively straightforward. The sufficiency, however, is not. In the proof for the sufficiency of conditions (i) and (ii), the authors of \cite{LiuStructural} intended to show that if the switched system (\ref{plant-switched})
						is not structurally controllable and condition (i) holds, then
						condition (ii) cannot hold, i.e., ${\rm grank}[A_1,...,A_N, B_1,...,B_N]<n$. To achieve this, the authors argued that if for every matrix pair $(\tilde {\bar A},\tilde {\bar B})$, where $\tilde {\bar A}=\sum \nolimits_{i=1}^N \bar u_i\bar A_i$, $\tilde {\bar B}=\sum \nolimits_{i=1}^m \bar u_i\bar B_i$, $(\bar A_i, \bar B_i)$ can be any realization of $(A_i,B_i)$, and $\bar u_1,...,\bar u_N\in {\mathbb R}$, there is a nonzero vector $q$ such that $q\tilde {\bar A}=0$ and $q\tilde {\bar B}=0$, then ${\rm grank}[A_1,...,A_m, B_1,...,B_m]=n$ cannot hold. However, this claim is not necessarily true. The following counter-example demonstrates this.
						
						\begin{example} \label{example1}
							Consider a switched system with $N=2$, whose subsystem parameters are ($a_{21},a_{31},b_1\in {\mathbb R}$):
							$$A_1=\left[ \begin{array}{ccc}
								0 & 0 & 0 \\
								a_{21} & 0 & 0 \\
								0 & 0 & 0 \\
							\end{array}\right], B_1=\left[\begin{array}{c}
								b_1\\
								0\\
								0
							\end{array}\right],A_2=\left[\begin{array}{ccc}
								0 & 0 & 0 \\
								0 & 0 & 0 \\
								a_{31} & 0 & 0 \\
							\end{array}\right],$$and $B_2=[0;0;0]$.
							{Notice that condition (i) is satisfied for this system and ${\rm grank}[A_1,A_2,B_1,B_2]=3$}. However, for $\bar u_1, \bar u_2\in {\mathbb R}$,
							$$[\bar u_1A_1+\bar u_2A_2, \bar u_1B_1+\bar u_2B_2 ]=\left[\begin{array}{cccc}
								0 & 0 & 0 & \bar u_1b_1 \\
								\bar u_1a_{21} & 0 & 0 & 0\\
								\bar u_2a_{31} & 0 & 0 & 0 \\
							\end{array} \right].$$Obviously, for any values of $a_{21},a_{31},b_1$ and $\bar u_1, \bar u_2$,
							${\rm rank}[\bar u_1A_1+\bar u_2A_2, \bar u_1B_1+\bar u_2B_2]< 3$. Hence, there exists a corresponding nonzero vector $q$, such that
							$q[\bar u_1A_1+\bar u_2A_2, \bar u_1B_1+\bar u_2B_2]=0$.
						\end{example}
						
						In fact, the essential idea of the above-mentioned proof in \cite{LiuStructural} is to derive the structural uncontrollability of the switched system $(A_i,B_i)|_{i=1}^N$ from the structural uncontrollability of the LTI system $(\tilde {\bar A},\tilde {\bar B})$. However, this direction is not necessarily true, since the structural controllability of $(A_i,B_i)|_{i=1}^N$ does not imply the structural controllability of $(\tilde {\bar A},\tilde {\bar B})$. The primary goal of this paper is to develop new criteria for
						structural controllability of switched systems, and meanwhile, rigorously establish the sufficiency (and necessity) of conditions (i) and~(ii).

						\begin{figure}
							\centering
							\includegraphics[width=3.20in]{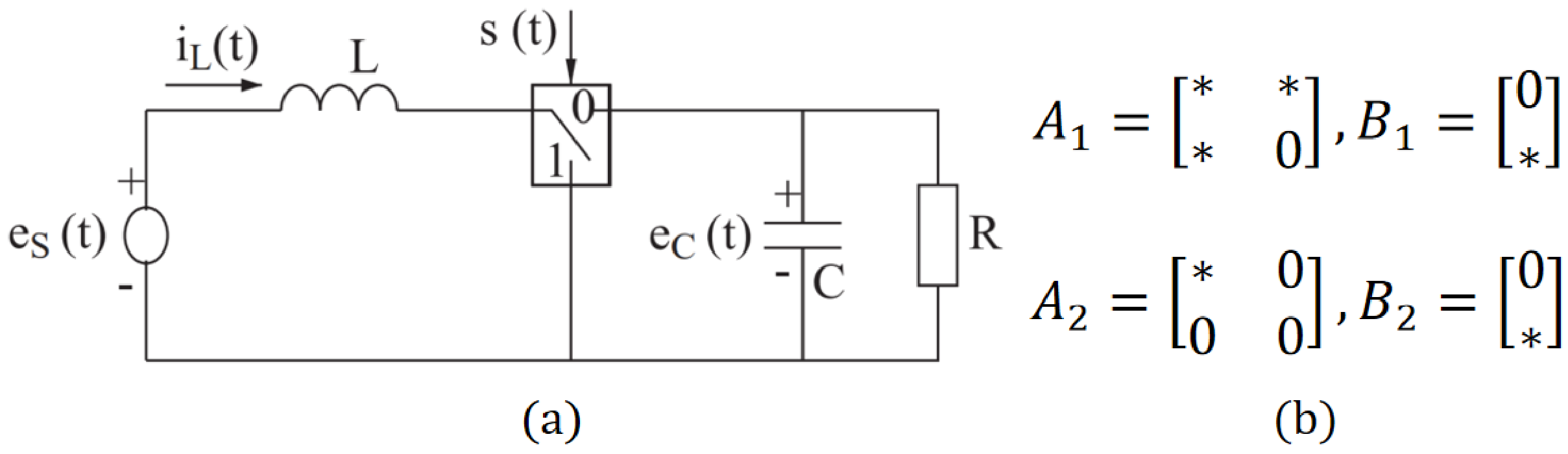}\\
							\caption{(a): a boost converter (borrowed from \cite{LiuStructural}). (b): the corresponding structured system model of (a), where $*$ represents nonzero entries and $0$ represents zero entries. }\label{switched-example}
						\end{figure}
						
						%However, we declare that Theorem 9 is still right. We shall provide a complete proof for this claim in the following.
						
						\section{Main Results}\label{sec-III} %Generalized cactus configuration and structural controllability
						This section presents a novel generalized cactus based criterion for the structural controllability of switched systems. Based on it, the sufficiency of conditions (i) and (ii) is established. Key to this new criterion is the introduction of some novel graph-theoretic concepts, detailed in the first two subsections.
						\subsection{Multi-layer dynamic graph}
						{Our basic idea is to analyze the rank of ${\cal R}$. Notice that there are $(nN)^n$ column blocks in ${\mathcal R}$. Inspired by \cite{Z.S2002Controllability}, we first develop a recursive expression of ${\rm span}{\cal R}$, which benefits mapping subcolumns of ${\cal R}$ to directed graphs. Below, bold capital letters denote subspaces.}
						
						Given a subspace ${\bf V}\subseteq {\mathbb R}^n$, let $\Gamma_{A}{\bf V}=\sum_{i=0}^{n-1}A^i{\bf V}$.
						Let $\left \langle A|B\right \rangle=\sum_{i=0}^{n-1}A^i{\rm Im} B$ be the controllable subspace of $(A,B)$, with ${\rm Im} B$ the subspace spanned by the columns of $B$. From \cite{Z.S2002Controllability}, the controllable subspace of system (\ref{plant-switched}), denoted by $\bf \Omega$, can be iteratively expressed as
						${\mathbf{\Omega}}_1=\sum\nolimits_{i=1}^{N}\langle A_i|B_i \rangle$, and
							${\mathbf{\Omega}}_i=\sum\nolimits_{k=1}^N\Gamma_{A_{k}} {\bf \Omega}_{i-1}$, for $i=2,...,n$.
 Then, ${\mathbf{\Omega}}={\rm span} {\cal R}={\mathbf{\Omega}}_n$.	Based on $\{{\bf \Omega}_i|_{i=1}^n\}$, {we recursively define a collection of subspaces $\{{\bf \Phi}_i|_{i=0}^{\infty}\}$ as}
						$$\begin{array}{c}
							{\bf \Phi}_{0}={\rm Im}\ B_1+{\rm Im} \ B_2+\cdots+ {\rm Im} \ B_N,\\
							{\bf \Phi}_{j}=\sum \nolimits_{i=1}^N A_i{\bf \Phi}_{j-1}, j=1,....\infty
						\end{array}
						$$
						Construct $\{{\bf W}_j|_{j=-1}^{\infty}\}$ as ${\bf W}_{-1}=\emptyset$, ${\bf W}_j={\bf W}_{j-1}+{\bf \Phi}_j$ for $j=0,1,\cdots, \infty$. This implies ${\bf W}_j=\sum_{i=0}^j{\bf \Phi}_i$ ($j=0,1,\cdots, \infty$), ${\bf W}_0\subseteq {\bf W}_1\subseteq \cdots \subseteq {\bf W}_{\infty}$, and ${\bf W}_{\infty}={\bf  \Omega}$. It turns out that ${\bf W}_j={\bf W}_{j-1}+\sum \nolimits_{i=1}^N A_i{\bf W}_{j-1}$. Therefore, if for some $j$, ${\bf W}_j={\bf W}_{j+1}$, then ${\bf W}_{j+2}={\bf W}_{j+1}+\sum \nolimits_{i=1}^N A_i{\bf W}_{j+1}={\bf W}_{j}+\sum \nolimits_{i=1}^N A_i{\bf W}_{j}={\bf W}_{j+1}$, leading to ${\bf W}_{k}={\bf W}_{j}={\bf  \Omega}$ for any $k\ge j$. This means there exists some $l_0\le n$, such that ${\bf W}_{\bar l}={\bf  \Omega}$ for any $\bar l\ge l_0$. %It is worth noting that the difference between the subspaces $\{{\bf \Phi}_j|_{j=0}^{\infty}\}$ and those in \citep[Sec 4.3]{Z.S2002Controllability} lies in that, ${\bf \Phi}_j$ does not necessarily contain ${\bf \Phi}_{j-1}$, which %-{\rm dim} {\bf \Phi}_0
						%is important to reduce the redundant edges in constructing the associated dynamic graphs shown below.
						Corresponding to $\{{\bf \Phi}_i|_{i=0}^{\infty}\}$, define the matrix series $\{\Gamma_j|_{j=0}^{\infty}\}$ as
						$$\Gamma_{0}=[B_1,...,B_N],\Gamma_{j}=[A_i\Gamma_{j-1}|_{i=1}^N], j=1,2...$$
						{It follows that ${\bf \Phi}_i={\rm span}\Gamma_i$.} Moreover, let
						$$W_0=\Gamma_{0},W_j=[W_{j-1},\Gamma_j],j=1,...\infty$${We have ${\bf W}_j={\rm span}W_j$, $j=1,...,\infty$.}
						Following the above analysis on $\{{\bf W}_j|_{j=-1}^{\infty}\}$,  if ${\rm rank} W_j= {\rm rank} W_{j+1}$ for some $j$, then ${\rm rank} W_k= {\rm rank} W_j={\rm rank} {\cal R}$ for any $k\ge j$.
						Since all these relations hold for any numerical matrices, they must hold when `${\rm rank}$' is replaced by `${\rm grank}$' (the corresponding matrices $(A_i,B_i)$ become structured ones). {It can be seen that $W_k$ consists of all distinct items $A_{i_n}^{j_n}\cdots A_{i_1}^{j_1}B_{i_1}$ in (\ref{control-m}) such that $j_n+\cdots +j_1\le k$, $1\le i_1\le N$.}
						
						%Let $l_0=n-{\rm rank} \Gamma_{0}$. Then, ${\rm rank}{\mathcal R}= {\rm rank} W_{n-1} = {\rm rank} W_{l_0}$.

						 %, which extends the dynamic graph adopted in \cite{poljak1990generic} for LTI systems to switched systems
						%	{\linespread{1.15} \selectfont %(in a way analogous to that $W_i$ is obtained by incorporating $W_{i-1}$ and $\Gamma_i$; the precise relation between $\hat {\cal G}_i$ and $W_i$ will be explained subsequently)
							Inspired by \cite{poljak1990generic}, we construct the dynamic graphs $\{\hat {\cal G}_i|_{i=0}^{\bar l}\}$ associated with $\{W_i|_{i=0}^{\bar l}\}$ iteratively, with $\bar l\ge l_0\doteq n-{\rm grank}\Gamma_{0}$. {The relationship between $\hat {\cal G}_i$ and $W_i$ will be explained subsequently.} At the beginning,  ${\hat {\cal  G}}_0=(\hat V_0, \hat E_0)$ with $\hat V_0= \hat U_{0}\cup \hat X_{0}$, $\hat X_{0}=\{x_{j0}^{00}:j=1,...,n\}$, $\hat U_{0}=\{u_{k,j0}^{00}: j=1,...,N,k=1,...,m_j\}$, and $\hat E_0=\{(u_{k,j0}^{00},x_{q0}^{00}): B_{j,qk}\ne 0, j=1,...,N, k=1,...,m_j\}$. For $i\ge 1$,  ${\hat {\cal G}}_i$ is obtained by adding vertices $\hat V_i=\hat U_{i}\cup \hat X_{i}$ and the associated edges $\hat E_i\cup \hat E_{i,i-1}$ to $\hat {\cal G}_{i-1}$ {. We shall call the subgraph of ${\hat {\cal G}}_{i}$ induced by $\hat V_i$ the $i$th layer.
							 Here, $\hat X_i$ consists of $N^{i-1}$ copies of $X_k$ for each $k\in [N]$, and denote the $t$th copy of the $j$th vertex in $X_k$ by $x_{ji}^{kt}$, yielding $\hat X_{i}=\{x_{ji}^{kt}: j=1,...,n,k=1,...,N,t=1,...,N^{i-1}\}$.
							  Similarly, $\hat U_i$ consists of $N^{i}$ copies of $U_l$ for each $l\in [N]$, and denote $u^{kt}_{j,li}$ as the $(k,t)$th copy of the $j$th input vertex in $U_l$, yielding $\hat U_{i}=\{u^{kt}_{j,li}:k,l=1,...,N,j=1,...,m_l,t=1,...,N^{i-1}\}$ (notice that a pair $(k,t)$ is used to index the copies).} In addition, the edge set $\hat E_i=\{(u^{kt}_{j,li},x^{kt}_{qi}): k,l=1,...,N, t=1,...,N^{i-1}, j=1,...,m_l, B_{l,qj}\ne 0\}$, i.e., the $t$th ($\forall t\in [N^{i-1}]$) copy of $X_k$ ($\forall k\in [N]$) is connected to the $(1,t)$th,...,$(N,t)$th copies of $U_{l}$ by edges corresponding to nonzero entries of $B_{l}$ ($\forall l\in [N]$).  The edge set
							 $\hat E_{i,i-1}$ is defined as
							  									$$\hat E_{i,i-1}\!=\!\left\{(x_{ji}^{kt},x_{p,i-1}^{k't'})\Bigg|\begin{array}{c}
							  	A_{k,pj}\ne 0, k=1,...,N,k'=1,...,N \\
							  	t'\!=\!1,...,N^{i-2},
							  	t\!=\!(k'-1)N^{i-2}+t'
							  \end{array} \right\},$$
						  i.e.,  $\hat E_{i,i-1}$ consists of edges by connecting the $t'$th ($t'\in [N^{i-2}]$) copy of $X_{k'}$ ($k'\in [N]$) in $\hat X_{i-1}$ with the $(k'-1)N^{i-2}+t'$th copy of $X_k$ ($k\in [N]$) corresponding to nonzero entries of $A_k$. Here, we set $N^{-1}=0$, $x^{q0}_{p0}\equiv x^{00}_{p0}$ for $q=1,...,N$, and multiple edges are disallowed (i.e., $\hat E_{1,0}=\{(x_{j1}^{k1},x^{00}_{p0}):k=1,...,N, A_{k,pj}\ne 0\}$).  For an edge $e=(x_{ji}^{kt},x_{p,i-1}^{k't'})\in \hat E_{i,i-1}$, the weight $w(e)=A_{k,pj}$, and for an edge $e=(u^{kt}_{j,li},x^{kt}_{qi})\in \hat E_i$, its weight $w(e)=B_{l,qj}$. {See Fig. \ref{dynamic-graph} for an illustration of the construction process. As can be seen, in the state vertex notation $x_{ji}^{kt}$, superscripts $k$, $t$ denote the subsystem and copy indices, while subscripts $j,i$ indicate its order in $X_k$ and the layer index, respectively. Similarly, in $u_{j,li}^{kt}$, $(k,t)$ indices the copies, $l$ indices the subsystems, and $j,i$ indicate the order in $U_l$ and the layer index, respectively. Moreover, in the $i$th layer ($i\ge 1$), we copy the state vertex set $X_k$ of the $k$th subsystem $N^{i-1}$ times (per $k\in [N]$), and each copy of them is connected to a copy of input vertices of the $1,...,N$-th subsystems. Between two successive layers, there are edges (corresponding to nonzero entries of $A_k$, per $k\in [N]$) from each copy of state vertex set $X_k$ in $i$th layer ($N^{i-1}$ copies in total) to {\emph{one and only one}} copy of state vertex set of the $1,2,...,N$th subsystems in the $(i-1)$th layer, $k=1,...,N$, $i=1,...,\bar l$ (one can compare this process to the process that $W_i$ is obtained by incorporating $W_{i-1}$ and $[A_1\Gamma_{i-1},\cdots, A_N\Gamma_{i-1}]$).} Since $\hat {\cal G}_{\bar l}$ has multiple layers, we call it the {\emph{multi-layer dynamic graph}} (MDG). It can be seen that MDG generalizes the dynamic graph adopted in \cite{poljak1990generic} for LTI systems to switched systems.  %It can be seen that MDG generalizes the dynamic graph adopted in \cite{poljak1990generic} for LTI systems to switched systems.

						\begin{figure}
							\centering
							\includegraphics[width=3.35in]{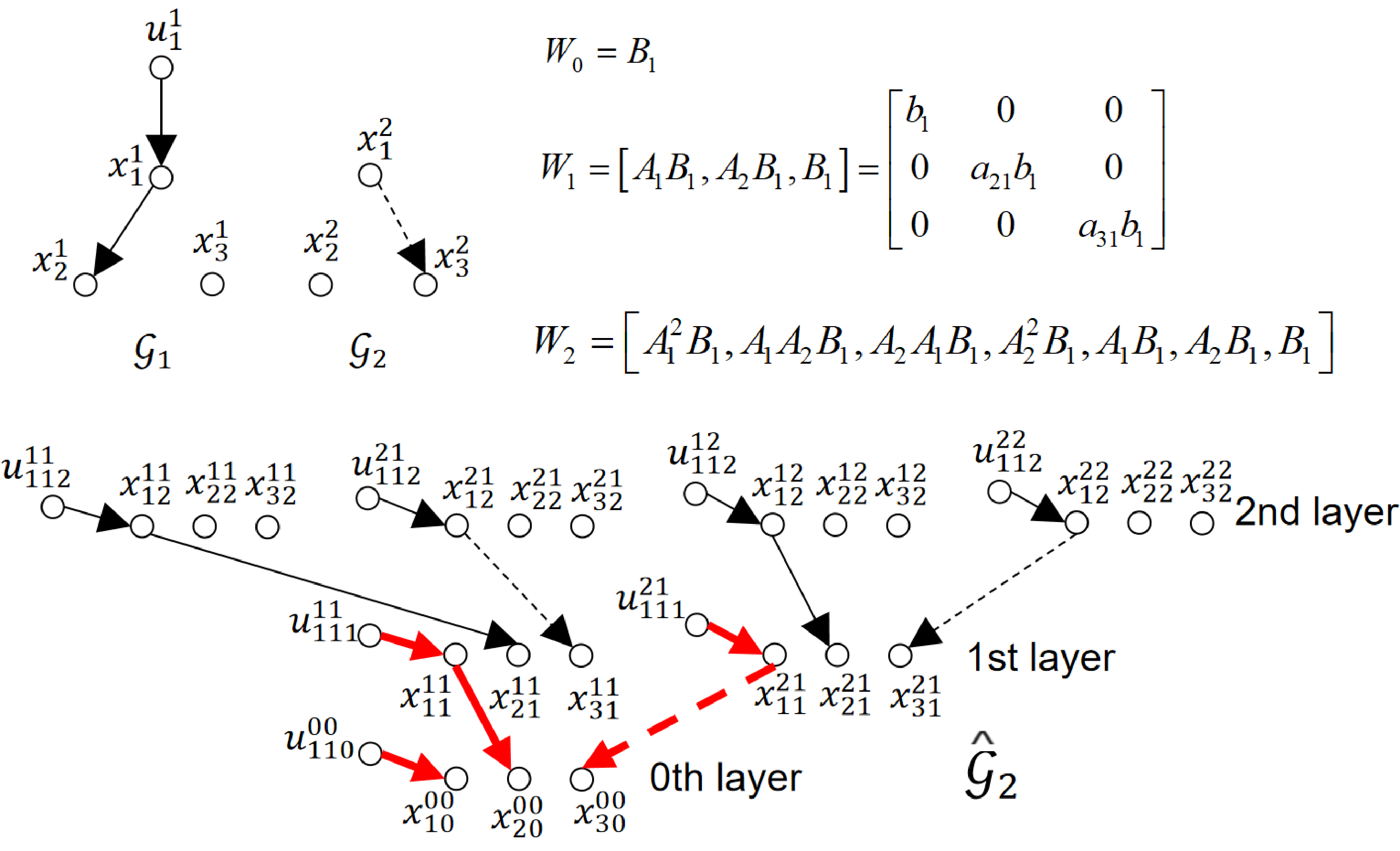}\\
							\caption{{Illustration of graph representations of a switched system with two subsystems (hereafter, edges with different color indices are depicted with different styles of lines). The subsystem digraphs ${\cal G}_1$ and ${\cal G}_2$ correspond to respectively $(A_1,B_1)$ and $(A_2,B_2)$ given in Example \ref{example1}. Next to ${\cal G}_1$ and ${\cal G}_2$ are $W_0,W_1$, and $W_2$. Graph $\hat {\cal G}_2$ is the MDG with $3$ layers.  Bold red lines represent a $\hat U_{}-\hat X_{0}$ linking $L$ with size $3$, which corresponds to $3$ nonzero product terms in $W_1=[B_1,A_1B_1,A_2B_1]$ that are in different rows and columns. The permutation $\pi$ associated with $L$ is $\pi(x_{10}^{00})=u_{110}^{00}$, $\pi(x_{20}^{00})=u^{11}_{111}$, and $\pi(x_{30}^{00})=u_{111}^{21}$. According to (\ref{determinant-2}), the determinant of $W_2({\rm head}(L), {\rm tail}(L))$ is $w(L)=a_{21}a_{31}b_1^3$.} }\label{dynamic-graph}
						\end{figure}

						A collection of $k$ vertex-disjoint paths $L=(p_1,...,p_k)$  in $\hat {\cal G}_{\bar l}$ is called a linking (with size $k$). We call $L$ a $S-T$ linking, if ${\rm tail}(L)\doteq \{{\rm tail}(p_i):i=1,...,k\}\subseteq S$, and ${\rm head}(L)\doteq \{{\rm head}(p_i): i=1,...,k\}\subseteq T$. We are interested in the linkings from $\hat U$ to $\hat X_{0}$, where $\hat U= \bigcup_{i=0}^{\bar l}\hat U_{i}$ (the dependence of $\hat U$ on $\bar l$ is omitted for simplification), i.e., $S-T$ linkings with $S\subseteq \hat U$ and $T\subseteq \hat X_{0}$. For such a linking $L=(p_1,...,p_k)$, define $w(p_i)$ as the product of weights of all individual edges in each path $p_i$ of $L$, $i=1,...,k$, and $w(L)=w(p_1)w(p_2)\cdots w(p_k)$.
						Let $\prec$ be a fixed order of vertices in $\hat U$ and $\hat X_{0}$.  For a linking $L=(p_1,...,p_k)$, let $s_1\prec s_2\prec\cdots \prec s_k$ and $t_1\prec t_2\prec\cdots \prec t_k$ be respectively
						the tails and heads of $L$. Moreover, suppose $s_{\pi{(i)}}$ and $t_{i}$ are the tail and head of the path $p_i$, $i=1,...,k$. Then, we obtain a permutation $\pi\doteq (\pi(1),...,\pi(k))$ of $(1,2,...,k)$, and denote its sign by ${\rm sign}(\pi)\in \{1,-1\}$. Recall that the sign of a permutation is defined as $(-1)^q$, where $q$ is the number of transpositions required to transform the permutation into an ordered sequence. {See Fig. \ref{dynamic-graph} for the illustration of a linking.}
						
						% For a linking $L=(p_1,...,p_k)$, let $s_1\prec s_2\prec\cdots \prec s_k$ and $t_1\prec t_2\prec\cdots \prec t_k$ be respectively
						%the tails and heads of $L$. Moreover, suppose $s_{\pi{(i)}}$ and $t_{i}$ are the tail and head of the path $p_i$, $i=1,...,k$. Then, we obtain a permutation $\pi\doteq (\pi(1),...,\pi(k))$ of $1,2,...,k$, and denote its sign by ${\rm sign}(\pi)\in \{1,-1\}$ (the sign of a permutation is defined in matrix determinant). The sign of the linking $L$ is defined as ${\rm sign}(L)={\rm sign}(\pi)$.
						%To illustrate the above construction, we provide next an example.	

						%We call such a nonzero term a product term in $W_{\bar l}$.
					{The MGD ${\hat {\cal G}}_{\bar l}$ and matrix $W_{\bar l}$ is related as follows.\footnote{Note that these relations hold between $\hat {\cal G}_{l}$ and $W_l$ for any $l\ge 0$.}} Let $A_i(j,p)$ be the $(j,p)$th entry of $A_i$, and $B_{i}(j,p)$ defined similarly. Consider a column block of $W_{\bar l}$, given by $A_{i_k}A_{i_{k-1}}\cdots A_{i_1}B_{i_0}$ ($k\le \bar l$). Observe that the $(t_k,t_0')$th entry of $A_{i_k}A_{i_{k-1}}\cdots A_{i_1}B_{i_0}$ is the sum of all {\emph{product terms}} in the form of
						\begin{equation}\label{sum-item}A_{i_k}(t_k,t_{k-1})A_{i_{k-1}}(t_{k-1},t_{k-2})\cdots A_{i_1}(t_1,t_0)B_{i_0}(t_0,t_0'),\end{equation}where $t_k,t_{k-1},...,t_0\in [n]$, $t_0'\in [m_{i_0}]$, $i_k,...,i_0\in [N]$, and each involved entry in the product term is nonzero.
						By the construction of $\hat {\cal G}_{\bar l}$, for each product term (\ref{sum-item}), there exists a path
						{\small{\begin{equation}\label{path-item} p\doteq (u_{t_0',i_0k}^{i_1,j_1},x^{i_1,j_1}_{t_0,k},x_{t_1,k-1}^{i_2,j_2},\cdots, x_{t_{k-1},1}^{i_{k},j_k},x^{00}_{t_k,0})\end{equation}}}with head in $\hat X_{0}$ and tail in $\hat U$ of $\hat {\cal G}_{\bar l}$, where $j_1,...,j_k$ are the corresponding copy indices in the $k$th,$\cdots, 1$st layers, respectively; and vice versa, that is, any $\hat U-\hat X_{0}$ path, say $p$, corresponds to a {\emph{unique}} product term in $W_{\bar l}$ in the form of (\ref{sum-item}). {Moreover, there are one-one correspondences between rows of $W_{\bar l}$ and the set $\hat X_{0}$, and between columns of $W_{\bar l}$ and the set $\hat U$. Take the switched system in Fig. \ref{dynamic-graph} for instance, where $(A_i,B_i)|_{i=1}^2$ are given in Example \ref{example1}. The columns of $W_1=[B_1,A_1B_1,A_2B_1]$ correspond to vertices $\{u_{110}^{00},u_{111}^{11},u_{111}^{21}\}$, and rows correspond to $\{x_{10}^{00},x_{20}^{00},x_{30}^{00}\}$. The path $(u_{111}^{21},x_{11}^{21},x_{30}^{00})$ corresponds to the product term $a_{31}b_1$. On the other hand, since there is no path from $\{u^{21}_{112}\}$ to $\hat X_0$ in $\hat {\cal G}_2$, all entries of $A_1A_2B_1$ are zero.}
						
						Denote by $W_{\bar l}(I,J)$ the submatrix of $W_{\bar l}$ given by rows corresponding to $I$ and columns corresponding to $J$, where $I\subseteq \hat X_0, J\subseteq \hat U$. Based on the above relations and inspired by \cite[Lem 1]{murota1990note}, the following proposition relates the determinant of a square submatrix $W_{\bar l}(I,J)$ to the $J-I$ linkings of~$\hat {\cal G}_{\bar l}$.
						
						\begin{proposition} \label{determinant}
							Let $W_{\bar l}(I,J)$ be a square submatrix of $W_{\bar l}$. Then
							\begin{equation} \label{linking-determinant}
								\det W_{\bar l}(I,J)=\sum_{L: \ J-I \ {\rm linking \ with \ size} \ |I|   \ {\rm in} \ \hat {\cal G}_{\bar l}} {\rm sign}(L)w(L),
							\end{equation}
							where $I\subseteq \hat X_0, J\subseteq \hat U$ and $|I|=|J|$.
						\end{proposition}
						
						\begin{proof}
						From (\ref{sum-item}), the entry at the position $(x_{t_k,0}^{00},u_{t_0',t_0k}^{i_1,j_1})$ of $W_{\bar l}$, with $x_{t_k,0}^{00}\in \hat X_{0}$, $u_{t_0',t_0k}^{i_1,j_1}\in \hat U$, is expressed as
							\begin{equation}\label{entry-expression}
								W_{\bar l}(x_{t_k,0}^{00},u_{t_0',t_0k}^{i_1,j_1})=\sum\nolimits_{p} w(p),
							\end{equation}where the summation is taken over all paths $p$ in $\hat {\cal G}_{\bar l}$ from $u_{t_0',t_0k}^{i_1,j_1}$ to $x_{t_k,0}^{00}$, i.e., all paths in the form of (\ref{path-item}).
							With the understanding of $J\subseteq \hat U$ and $I\subseteq \hat X_{0}$, by submitting (\ref{entry-expression}) into the expression of the determinant, we have
							({see Fig. \ref{dynamic-graph} for illustration})
							\begin{equation}\label{determinant-2} \det W_{\bar l}(I,J)=\sum\nolimits_{\pi} {\rm sign}(\pi) \prod\nolimits_{x_{t_k,0}^{00}\in I}W_{\bar l}(x_{t_k,0}^{00},\pi(x_{t_k,0}^{00})),\end{equation}
where the summation is taken over all permatations $\pi: I\to J$ such that $\prod\nolimits_{x_{t_k,0}^{00}\in I}W_{\bar l}(x_{t_k,0}^{00},\pi(x_{t_k,0}^{00}))\ne 0$.
							
							Notice that each nonzero $\prod\nolimits_{x_{t_k,0}^{00}\in I}W_{\bar l}(x_{t_k,0}^{00},\pi(x_{t_k,0}^{00}))$ is the product of weights of $|I|$ paths, each path from $\pi (i)\in J$ to $i\in I$, given by $p_{\pi(i),i}$. Let $P_{\pi}$ be the collection of such $|I|$ paths. If two paths $p_{\pi(i),i}, p_{\pi(i'),i'}\in P_{\pi}$ intersect at a vertex $v$, let $w$ and $\sigma$ be respectively the path from $\pi(i)$ to $v$ and the path from $v$ to $i$ in $p_{\pi(i),i}$, and $w'$ and $\sigma'$ be the path from $\pi(i')$ to $v$ and
							the path from $v$ to $i'$ in $p_{\pi(i'),i'}$. Then, two new paths are constructed by connecting $w$ with $\sigma'$ and $w'$ with $\sigma$, and the remaining paths remain unchanged. This produces a new collection of $|I|$ paths, denoted by $P_{\pi'}$, with $\pi'(i)=\pi(i')$ and $\pi'(i')=\pi(i)$, leading to ${\rm sign}(\pi')=-{\rm sign}(\pi)$, while $\prod \nolimits_{p_{i}\in P_{\pi}} w(p_{i})=\prod \nolimits_{p_{i}\in P_{\pi'}} w(p_{i})$. It is easy to see the correspondence between such $P_{\pi}$ and $P_{\pi'}$ is one-to-one in all $|I|$ paths from $J$ to $I$. Consequently, all collections of $|I|$ paths from $J$ to $I$ that are not vertex-disjoint will cancel out in (\ref{determinant-2}). This leads to (\ref{linking-determinant}).
						\end{proof}
						
						\begin{remark}
							A direct corollary of Proposition \ref{determinant} is that the generic dimension of controllable subspaces ${\rm grank}{\cal R}$ is no more than the maximum size of a $\hat U-\hat X_{0}$ linking in $\hat {\cal G}_{\bar l}$, $\bar l\ge l_0$.
						\end{remark}
						
						\subsection{Generalized stem, bud, and  cactus walking/configuration}
						Now we introduce some new graph-theoretic notions, namely, {\emph{generalized stem, generalized bud, generalized cactus configuration, and generalized cactus walking}}, which extends the corresponding graph-theoretic concepts from LTI systems to switched systems. These extensions are crucial to our results.
						
						\begin{definition}[Generalized stem] \label{def-general-stem}
							A subgraph ${\cal G}_s\!=\!(V_s,E_s)$ of the color union graph ${\cal G}_c$ is said to be a generalized stem, if it satisfies:
							\begin{itemize}
								\item There is only one input vertex $u\in V_s$ and no cycle;
								\item Each state vertex $x\in V_s\backslash \{u\}$ has exactly one ingoing edge (thus $|E_s|=|V_s|-1$);
								\item All edges of $E_s$ are S-disjoint.
							\end{itemize}
						\end{definition}

						\begin{definition}[Generalized bud] \label{def-general-bud}
							A subgraph ${\cal G}_s=(V_s,E_s)$ of the color union graph ${\cal G}_c$ is said to be a generalized bud, if ${\cal G}_s$ satisfies:
							\begin{itemize}
								\item There is no input vertex in $V_s$ and only one cycle;
								\item Every state vertex $x\in V_s$ is input-reachable in ${\cal G}_c$;
								\item Each state vertex $x\in V_s$ has exactly one ingoing edge (thus $|E_s|=|V_s|$);
								\item All edges of $E_s$ are S-disjoint.
							\end{itemize}
						\end{definition}
						
						Fig. \ref{example-cactus} presents examples of generalized stems and buds {contained in a ${\cal G}_c$}.  It can be verified that when ${\cal G}_c$ is the system digraph of an LTI system (i.e., without being colored), the generalized stem collapses to a conventional stem, and the generalized bud collapses to a cycle which is input-reachable.\footnote{It should be mentioned that in the original definition, a bud includes an edge that connects a cycle with a vertex out of this cycle \cite{C.T.1974Structural}. We do not include this edge in extending this definition to simplify descriptions.}
						
						\begin{definition} \label{cacuts-config}
							A subset $X_s\subset X$ is said to be covered by a generalized cactus configuration, if there is a collection of vertex-disjoint {generalized} stems and {generalized} buds that cover $X_s$.
						\end{definition}
						%	Below all vertices and edges belong to ${\cal G}_{c}$.

						%	\begin{example}
%							\begin{figure}
%								\centering
%								% Requires \usepackage{graphicx}
%								\includegraphics[width=1.3in]{stem-bud-example2.eps}\\
%								\caption{Example of a generalized stem (a) and generalized bud (b). In (a), `u' represents an input vertex. Note that there are $3$ different color indices, depicted by $3$ different line styles.}\label{example-stem-bud}
%							\end{figure}
							%	\end{example}
						
						\begin{figure}
							\centering
							\includegraphics[width=2.35in]{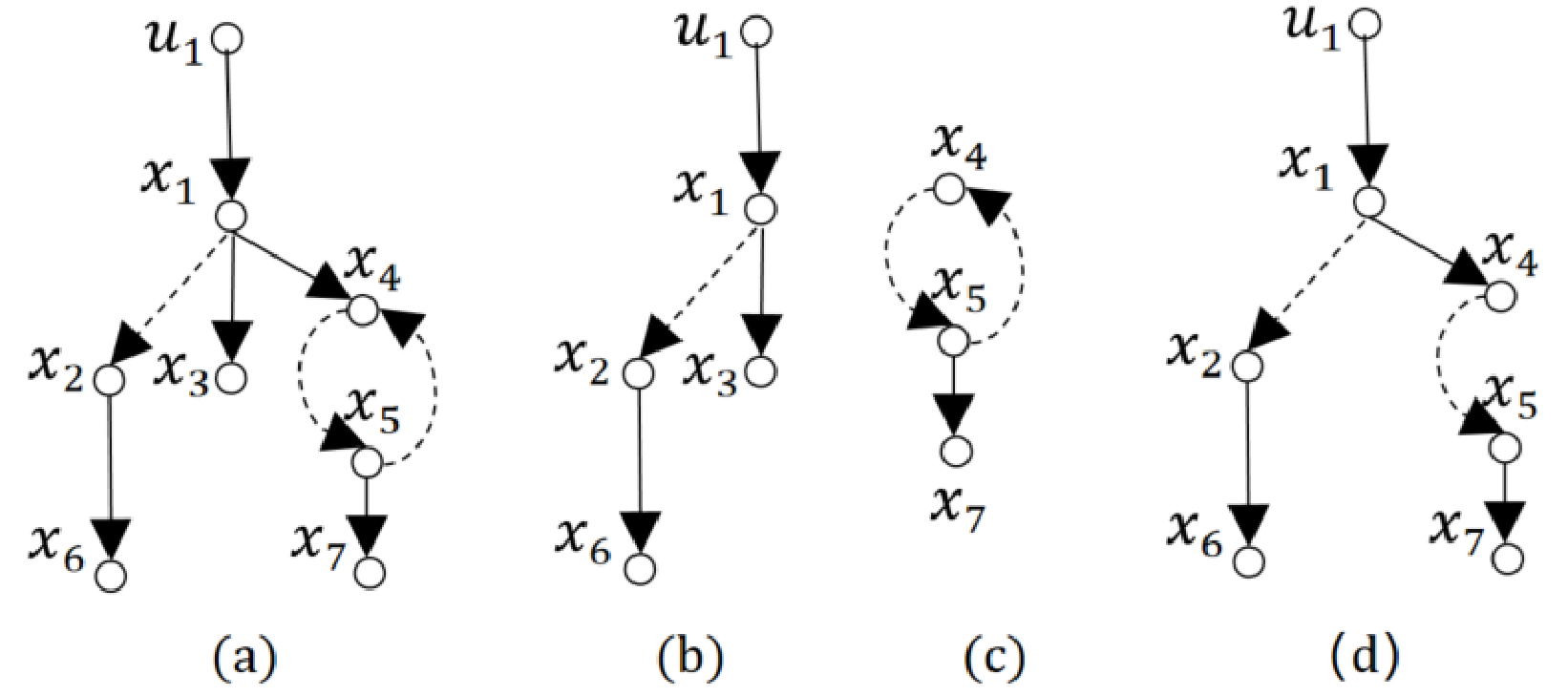}\\
							\caption{{(a): ${\cal G}_c$ of a switched system with $N=2$ and $n=7$. No multiple edges exist in ${\cal G}_c$. (b) and (d) are two generalized stems, and (c) is a generalized bud contained in ${\cal G}_c$.}}\label{example-cactus}
						\end{figure}
						
						Before introducing the notion {of} {\emph{generalized cactus walking}}, some concepts related to walks over the colored union graph ${\cal G}_{c}$ are presented.
						Given a walk $p=(e^{i_1}_{j_1,j_2},e^{i_2}_{j_2,j_3},...,e^{i_k}_{j_k,j_{k+1}})$ of ${\cal G}_{c}$, the {\emph{color index}} of $p$, given by ${\rm clo}(p)$, is the {sequence of color indices $(i_1,...,i_k)$ of edges $e^{i_1}_{j_1,j_2},...,e^{i_k}_{j_k,j_{k+1}}$}. For example, the walk $p=(e^3_{5,1},e^2_{1,2},e^2_{2,1},e^1_{1,3})$ has color index ${\rm clo}(p)=(3,2,2,1)$. {One can see that the color index of a walk is actually the sequence of subsystem indices indicating which subsystem the respective edge of this walk belongs to. Due to the S-disjointness of edges in a generalized stem or bud, let $p_1,...,p_k$ be $k$ ($k\ge 2$) walks starting from the {\emph{same}} vertex and ending at $k$ different vertices in a generalized stem or bud, then ${\rm col}(p_1),..., {\rm col}(p_k)$ are mutually different.}  {The {\emph{reverse edge}} of an edge $e\doteq(v_i,v_j)$ is the edge $e^c\doteq(v_j,v_i)$ by reversing its direction, $v_i\in U\cup X$, $v_j\in X$.} For a walk $p=(e_1,...,e_k)$, its {\emph{reverse walk}} is given by $p^c=(e^c_k,...,e^c_1)$, i.e., obtained by reversing the direction of this walk. Given two walks $p_1=(v_{i_1},...,v_{i_k})$ and $p_2=(v_{j_1},...,v_{j_{k'}})$, their {\emph{first intersection vertex}}, is defined to be ${\rm {ins}}(p_1,p_2)=v_{i_{k^*}}$ such that $k^*=\min\{q: i_q=j_{q}, q\in \{1,...,k\} \}$, i.e., being the first vertex they meet at when two walkers move along each individual walk simultaneously. Note that ${\rm {ins}}(p_1,p_2)={\rm {ins}}(p_2,p_1)$. It follows that if $p_1$ and $p_2$ are vertex disjoint, then ${\rm {ins}}(p_1,p_2)=\emptyset$. In addition, if ${\rm {ins}}(p_1,p_2)= v_{i_{k^*}}\ne \emptyset$, we use ${\rm Pins}(p_1\backslash p_2)$ to denote the subgraph of $p_1$ from $v_{i_1}$ to ${\rm {ins}}(p_1,p_2)$, i.e., ${\rm Pins}(p_1\backslash p_2)=(v_{i_1},v_{i_2},...,v_{i_{k^*}})$. Similarly, ${\rm Pins}(p_2\backslash p_1)=(v_{j_1},v_{j_2},...,v_{j_{k^*}})$.
							% given any sequence of color indices starting from the input vertex, there is at most one path in it that may correspond to it.}
						
						An input-state walk is a walk with head in $X$ and tail in $U$ of ${\cal G}_c$. For an input-state walk $p=(e^{i_1}_{j_1,j_2},e^{i_2}_{j_2,j_3},...,e^{i_k}_{j_k,j_{k+1}})$, there is
						a unique walk $(u_{j_1,i_1,k-1}^{i_2,\bullet},x_{j_2,k-1}^{i_2,\bullet},x_{j_3,k-2}^{i_3,\bullet},...,x_{j_k,1}^{i_k,1},x_{j_{k+1},0}^{00})$ from $\hat U$ to $\hat X_{0}$ in ${\hat {\cal G}}_{\bar l}$, where $\bar l\ge |p|$, and $\bullet$ is the omitted copy indices ({from (\ref{sum-item}), this walk corresponds to the $(j_{k+1},j_1)$th entry of $A_{i_k}\cdots A_{i_2}B_{i_1}$ in $W_{\bar l}$}). We call such a path the {\emph{MDG-path of $p$}}, and use $\hat p$ to denote the MDG-path of $p$ (in the corresponding dynamic graph ${\hat {\cal G}}_{\bar l}$, $\bar l\ge |p|$). {\emph{All notations for $p$ (e.g., ${\rm Pins}(\cdot)$) are applicable to $\hat p$.}} Specially, for an edge $e=(x_{ji}^{ks},x_{p,i-1}^{k's'})\in \hat E_{i,i-1}$, its color index is~$k$.    %in particular, if $x_{j_{q+1}}$ is the head of the $q$th edge $e^{i_q}_{j_q,j_{q+1}}$, then the corresponding $x_{j_{q+1},k-q}^{i_{q+1},\bullet}$ is in the $(k-q)$th layer of $\hat G_{\bar l}$, $q\in [k]$
						
				%		{To better understand the graph-theoretic concepts introduced here, one should keep in mind several correspondences between the MDG $\hat {\cal G}_l$ and $W_{l}$ for any $l\ge 0$. First, the length of an MDG-path $\hat p$ equals the the highest layer index of its vertices plus one.   }
				
						%	It is not difficult to see that an input-state walk $p_i$ with length ${\rm len}(p_i)$ corresponds to a unique path from $\hat U$ to $\hat X_{0}$ in ${\hat {\cal G}}_{\bar l}$, $\bar l={\rm len}(p_i)-1$. All notations introduced for $p_i$ are also valid for such a path. Specially, for an edge $e=(v_{ji}^{ks},v_{p,i-1}^{k's'})\in \hat E_{i,i-1}$, its color index is $k$.
						
						%	Given two walk $p$ in ${\cal G}_c$, if its head $v_t={\rm head}(p)$ is contained by some loop, denote ${\rm laloop}(p)$ as the last loop with head and tail being $v_t$ when traversing $p$ (such a loop is called the $v_t-v_t$ last loop, if exists).
						For two walks $p_1$ and $p_2$, if ${\rm head}(p_1)={\rm tail}(p_2)$, $p_1\vee p_2$ denotes the walk obtained by appending $p_2$ to $p_1$. For notation simplicity, we write $(p_1\vee p_2)\vee p_3$ as $p_1\vee p_2 \vee p_3$. If ${\rm tail}(p)={\rm heap}(p)$, define
						$$p^{\wedge k}\doteq \underbrace{p\vee p\vee \cdots \vee p}_{k\ {\rm times}}.$$
						
						\begin{definition}\label{general-cactus}
							A collection of input-state walks $\{p_1,...,p_k\}$ is called a {\emph{generalized cactus walking}} (with size $k$), if the corresponding MDG-paths $\{\hat p_1,...,\hat p_k\}$ form a linking in the MDG $\hat {\cal G}_{\bar l}$, with $\bar l\ge \max\{|p_1|,...,|p_k|\}$. The head of a generalized cactus walking is the set of heads of its walks.
						\end{definition}

						\begin{remark}
							For an LTI system, say $(A_i,B_i)$, a set $X_{i}'\subseteq X_{i}$ is said to be covered by a cactus configuration \cite{C.T.1974Structural}, if (i) every vertex $v\in X_{i}'$ is input-reachable in ${\cal G}_i$, and (ii) $X_{i}'$ is covered by a collection of vertex-disjoint stems and cycles of ${\cal G}_i$. Suppose $X_{i}'$ is covered by a cactus configuration. As we shall show, this cactus configuration can naturally introduce $|X_{i}'|$ input-state walks that form a generalized cactus walking with size $|X_i'|$ \cite{murota1990note}. This is why the terminologies {\emph{generalized cactus configuration}} in Definition \ref{cacuts-config} and {\emph{generalized cactus walking}} in Definition \ref{general-cactus} are used.
						\end{remark}
						
						The following property of walks is useful to show the vertex-disjointness of their corresponding MDG-paths.
						
						\begin{lemma}\label{dif-color-pro}
							For any two input-state walks $p_i$ and $p_j$ ($i\ne j$), their corresponding MDG-paths $\hat p_i,\hat p_j$ are vertex-disjoint, if either $p_i$ and $p_j$ are vertex-disjoint, or ${\rm Pins}(p_i^c\backslash p_j^c)$ and ${\rm Pins}(p_j^c\backslash p_i^c)$ have distinct color indices.
						\end{lemma}
						
						\begin{proof}
							If $ p_i$ and $ p_j$ are vertex-disjoint in ${\cal G}_c$, then $\hat p_i$ and $\hat p_j$ are obviously so in ${\hat {\cal G}}_{\bar l}$ by definition. If ${\rm ins}(\hat p_i^c,\hat p_j^c)\ne \emptyset$, then we have ${\rm clo}({\rm Pins}(\hat p_i^c\backslash \hat p_j^c))= {\rm clo}({\rm Pins}(\hat p_j^c\backslash \hat p_i^c))$, which follows from the fact that all outgoing edges from the same copy of state vertex set are injected into the same copy of state vertex set in the lower layer. As a result, ${\rm clo}({\rm Pins}( p_i^c\backslash  p_j^c))= {\rm clo}({\rm Pins}( p_j^c\backslash  p_i^c))$,  contradicting the fact that ${\rm Pins}(p_i^c\backslash p_j^c)$ and ${\rm Pins}(p_j^c\backslash p_i^c)$ have distinct color indices.
						\end{proof}
						
						The following result reveals that each generalized stem (bud) can induce a generalized cactus walking, which is crucial for Theorem \ref{main-theorem}.
						\begin{proposition} \label{property-stem-bud}	
							The following statements are true:
							
							(1) If ${\cal G}_c$ contains a generalized stem that covers $X_s\subseteq X$, then there is a generalized cactus walking whose head is $X_s$.
							
							(2) If ${\cal G}_c$ contains a generalized bud that covers $X_s\subseteq X$, then there is a generalized cactus walking whose head is $X_s$, and the length of the shortest walk in it can be arbitrarily large.
							%	Suppose ${\cal G}_c$ contains a generalized stem that covers $X_s\subseteq X$. Then, there is a generalized cactus walking with size $|X|$. The same is true when the generalized stem is replaced with a generalized bud.
						\end{proposition}
	\begin{proof}
	For a generalized stem, denoted as ${\cal G}_{\rm stem}$, suppose it consists of vertices $u,x_1,...,x_k$, where $u$ is the unique input vertex. Since there is no cycle in ${\cal G}_{\rm stem}$ and each state vertex has only one ingoing edge, there is a unique path from $u$ to $x_1,...,x_k$ in ${\cal G}_{\rm stem}$, denoted by $p_{ux_1},...,p_{ux_k}$, respectively. We are to show that $\{p_{ux_1},...,p_{ux_k}\}$ is a generalized cactus walking. To this end, for any two $p_{ux_i}$ and $p_{ux_j}$ ($i\ne j$), consider a schedule of $2$ walkers along the paths $p_{ux_j}^c$ and $p_{ux_i}^c$ satisfying the following rules:
	\begin{itemize}
		\item At the time $t=0$, they are located at $x_j$ and $x_i$, respectively;
		\item The walker starting from $x_j$ (called walker $j$; similar for $x_i$) located at vertex $x_{k_1}$ at time $t$ must move along the path $p_{ux_j}^c$ (resp. $p_{ux_i}^c$) to a neighboring vertex $x_{k_2}$ such that $(x_{k_1}, x_{k_2})$ is an edge of $p_{ux_j}^c$ (resp. $p_{ux_i}^c$);
		\item In case a walker reaches $u$ at time $t$, it will leave $u$ at time $t+1$.
	\end{itemize}
	It can be seen that if the walker $j$ is located at vertex $x_{k_1}$ at time $t$, then the corresponding MDG-path $\hat p_{ux_j}$ will pass through $x^{\star,\bullet}_{k_1,t}$ for some $\star$ and $\bullet$. In this schedule, if $p_{ux_i}$ and $p_{ux_j}$ have different lengths, then walker $i$ and walker $j$ will not be located at the same position at the same time (then $|p_{ux_i}|=|p_{ux_j}|$), implying that $\hat p_{ux_i}$ and $\hat p_{ux_j}$ cannot intersect. Otherwise, if walker $i$ and walker $j$ are located at the same position ${\rm ins}(p_{ux_i}^c,p_{ux_j}^c)$ at the same time $t$ for the first time, the edges that they walk along from time $t-1$ to time $t$ must have different color indices (by the S-disjointness condition). This implies that ${\rm Pins}(p_{ux_i}^c\backslash p_{ux_j}^c)$ and ${\rm Pins}(p_{ux_j}^c\backslash p_{ux_i}^c)$ have distinct color indices. By Lemma \ref{dif-color-pro}, $\hat p_{ux_i}$ and $\hat p_{ux_j}$ are vertex-disjoint.
	
	Now consider a generalized bud, denoted by ${\cal G}_{\rm bud}$. Suppose it consists of vertices $x_1,...,x_k$, in which vertices $x_1,...,x_r$ form a cycle
	$p_{\rm cyc}\doteq (x_1,x_2,...,x_r,x_1)$, $1\le r\le k$. By definition, if we remove any one edge from $p_{\rm cyc}$, say, $(x_{i-1},x_{i})$ ($1\le i\le r$; if $r=1$, the edge becomes $(x_1,x_1)$), and add a virtual input $u'$ as well as the edge $(u',x_{i})$ to ${\cal G}_{\rm bud}$, then the obtained graph, given by ${\cal G}_{\rm bud\to stem}$,  becomes a generalized stem.
	Let $p_{x_ix_1},...,p_{x_ix_k}$ be the unique path from $x_i$ to $x_1,...,x_k$ in ${\cal G}_{\rm bud\to stem}$, respectively, where $p_{x_ix_i}=\emptyset$.
	From the above argument on the generalized stem, we know the corresponding MDG-paths $\hat p_{x_ix_1},...,\hat p_{x_ix_k}$ of $p_{x_ix_1},...,p_{x_ix_k}$ are vertex-disjoint. Let $p_{x_i\to x_i}$ be the cycle from $x_i$ to $x_i$ in ${\cal G}_{\rm bud}$, and let $p_{u\to x_i}$ be a path from an input vertex $u\in U$ to $x_i$ in ${\cal G}_c$. By definition, such a path always exists. For $q\in {\mathbb N}$, let
	$$\begin{array}{c}p^q_{ux_1}=p_{u\to x_i}\vee p_{x_i\to x_i}^{\wedge q}\vee p_{x_ix_1},\\
		\vdots\\
		p^q_{ux_k}=p_{u\to x_i}\vee p_{x_i\to x_i}^{\wedge q}\vee p_{x_ix_k}.\end{array}$$
	By adopting the walking schedule rule mentioned above, since $\hat p_{x_ix_1},...,\hat p_{x_ix_k}$ are vertex-disjoint, prefixing the same path to them will not affect the vertex-disjointness. Consequently, the collection of walks $\{p^q_{ux_1},...,p^q_{ux_k}\}$ corresponds to a collection of vertex-disjoint MDG-paths $\{\hat p^q_{ux_1},...,\hat p^q_{ux_k}\}$ with size $k$, for any $q\in {\mathbb N}$. Since $q$ can be arbitrarily large, the length of the shortest walk in $\{p^q_{ux_1},...,p^q_{ux_k}\}$ can be so, too.
\end{proof}
						
						\subsection{Criteria for structural controllability}
						Based on Proposition \ref{property-stem-bud}, the following theorem reveals that the existence of a generalized cactus configuration covering $X$ is sufficient for structural controllability.
						
						\begin{theorem} \label{main-theorem}
							If ${\cal G}_c$ contains a collection of vertex-disjoint generalized stems and generalized buds that cover $X$, then system (\ref{plant-switched}) is structurally controllable.
						\end{theorem}
						
						\begin{proof} We divide the proof into two steps. In the first step, we show a linking can be obtained from the union of generalized cactus walkings associated with a collection of vertex-disjoint generalized stems and buds. In the second step, we show that no other terms can cancel out the weight product term associated with this linking in~(\ref{linking-determinant}).
							
							%    		. In the second step, we prove that the specifically constructed generalized cactus walking corresponds to a linking in the associated MDG. In the third step, we show that no other terms can cancel out the
							%    	product associated with this linking. The details are as follows.

		Step 1: Suppose there are $s\in {\mathbb N}$ generalized stems, given by ${\cal G}_{\rm stem}^1,...,{\cal G}_{\rm stem}^{s}$, and $d\in {\mathbb N}$ generalized buds, given by ${\cal G}_{\rm bud}^1,...,{\cal G}_{\rm bud}^{d}$, all of them vertex-disjoint. Suppose the vertex set of ${\cal G}_{\rm bud}^i$ is $\{x_{i1},...,x_{id_i}\}$, for $i=1,...,d$ (hence $d_i$ is the number of vertices in ${\cal G}_{\rm bud}^i$), and for $i=1,...,s$, the vertex set of ${\cal G}_{\rm stem}^i$ is $\{u'_{i0},x'_{i1},...,x'_{is_i}\}$, in which $u'_{i0}$ is the unique input vertex (hence $s_i$ is the number of {\emph{state}} vertices in ${\cal G}_{\rm stem}^i$). Denote the unique path from $u'_{i0}$ to $x'_{ik}$ in ${\cal G}_{\rm stem}^i$ by $p_{u'_{i0}x'_{ik}}$, $i=1,...,s$, $k=1,...,s_i$.  It is not difficult to see that we can extend the unions of ${\cal G}_{\rm stem}^i|_{i=1}^s$ and ${\cal G}_{\rm bud}^i|_{i=1}^d$ to a subgraph ${\cal G}_{\rm cact}$ of ${\cal G}_{c}$ such that there is a {\rm unique} path without repeated vertices (in ${\cal G}_{\rm cact}$) from some input vertex $u_{i0}\in U$ to every state vertex of the cycle in ${\cal G}_{\rm bud}^i$, per $i\in \{1,...,d\}$.\footnote{If we regard each ${\cal G}_{\rm bud}^i$ and each ${\cal G}_{\rm stem}^i$ as a virtual node, then ${\cal G}_{\rm cact}$ becomes a directed tree (i.e., directed acyclic graph where each node has exactly one ingoing edge,  except for the roots that have no ingoing edges).} Without loss of generality, assume that $\{{\cal G}_{\rm bud}^i: i=1,...,d\}$ are partially ordered ${\cal G}_{\rm bud}^1 \preceq {\cal G}_{\rm bud}^2 \preceq \cdots \preceq  {\cal G}_{\rm bud}^d$ such that in ${\cal G}_{\rm cact}$, there is no edge starting from ${\cal G}_{\rm bud}^k$ to ${\cal G}_{\rm bud}^j$ if $k>j$. Moreover, assume that $x_{i1}$ is the head of the shortest path (in ${\cal G}_{\rm cact}$) from the input vertex $u_{i0}$ to a vertex of the cycle of ${\cal G}_{\rm bud}^i$, and denote by the path from $u_{i0}$ to $x_{i1}$ in ${\cal G}_{\rm cact}$ by $p_{u_{i0}\to x_{i1}}$, for $i=1,...,d$. The cycle from $x_{i1}$ to $x_{i1}$ in ${\cal G}_{\rm bud}^i$ is denoted by $p_{x_{i1}\to x_{i1}}$, and the shortest path from $x_{i1}$ to $x_{ik}$ is denoted by $p_{x_{i1}x_{ik}}$, $k=1,...,d_i$ (note $p_{x_{i1}x_{i1}}=\emptyset$). Then, construct a collection of walks $\{p^{q_i}_{u_{i0}x_{ik}}: i\in[d], k\in [d_i]\}$ as
\begin{equation}\label{construct} p^{q_i}_{u_{i0}x_{ik}}=p_{u_{i0}\to x_{i1}}\vee p_{x_{i1}\to x_{i1}}^{\wedge q_i}\vee p_{x_{i1}x_{ik}},\end{equation}
where $q_1,...,q_d$ are defined as follows:
\begin{align*}
	q_1&= \max\{s_1,...,s_s,2d^2_1\},\\
	q_2&= \max\{|p^{q_1}_{u_{10}x_{11}}|,...,|p^{q_1}_{u_{10}x_{1d_1}}|,2d_2^2\},\\
	q_{i}&=\max\{|p^{q_{i-1}}_{u_{{i-1},0}x_{{i-1},1}}|,...,|p^{q_{i-1}}_{u_{{i-1},0}x_{{i-1},d_{i-1}}}|,2d_i^2\}, i=2,...,d.
\end{align*}
We are to show that the collection of walks $\{p_{u_{i0}'x_{ik}'}: i=1,...,s, k=1,...,s_i\}\cup \{p^{q_i}_{u_{i0}x_{ik}}: i=1,...,d, k=1,...,d_i\}$ constructed above forms a generalized cactus walking. Due to the vertex-disjointness of ${\cal G}_{\rm stem}^i|_{i=1}^s$, it is obvious that the MDG-paths $\{\hat p_{u_{i0}'x_{ik}'}: i=1,...,s, k=1,...,s_i\}$ are vertex-disjoint. The vertex-disjointness of the MDG-paths $\{\hat p^{q_i}_{u_{i0}x_{ik}}: k=1,...,d_i\}$ within each generalized bud ${\cal G}_{\rm bud}^i$ has been demonstrated in Proposition \ref{property-stem-bud} for any $q_i\ge 0$. Given a $j\in\{1,...,d\}$, the vertex-disjointness between any path in $\{\hat p^{q_j}_{x_{j0}x_{jk}}: k=1,...,d_j\}$, say $\hat p_j$, and any one in $\{\hat p_{u_{i0}'x_{ik}'}: i=1,...,s, k=1,...,s_i\}\cup \{\hat p^{q_i}_{u_{i0}x_{ik}}: i=1,...,j-1, k=1,...,d_i\}$, say $\hat p_{j'}$, is demonstrated as follows. Observe that the number $q_j$ of repeated cycles $p_{x_{j1}\to x_{j1}}$ in $p_j$ is no less than $|p_{j'}|$ by the construction of $\{q_i:i=1,...,d\}$. As a result, each of the first $|p_{j'}|$ vertices in $p_j^c$ is different from any vertex in $p_{j'}^c$. Since $\hat p_{j'}^c$ terminates at the $(|\hat p_{j'}^c|-1)$th layer in the corresponding MDG, the last $|\hat p_{j}^c|-|\hat p_{j'}^c|$ vertices of $\hat p_{j}^c$ are in different layers from any vertex of $\hat p_{j'}^c$, which cannot intersect. Therefore, $\hat p_j$ and $\hat p_{j'}$ are vertex-disjoint. Taking together, we obtain that ${L}\doteq \{\hat p_{u_{i0}'x_{ik}'}: i=1,...,s, k=1,...,s_i\}\cup \{\hat p^{q_i}_{u_{i0}x_{ik}}: i=1,...,d, k=1,...,d_i\}$ is a linking with ${\rm head}(L)=\hat X_{0}$ in the MDG $\hat {\cal G}_{\bar l}$, where $\bar l$ is the largest length of a path in $L$.

Step 2: We are to demonstrate that ${\rm sign}(L)w(L)$ cannot be canceled out in (\ref{linking-determinant}), in which $I=\hat X_{0}$, $J={\rm tail}(L)$, and $W_{\bar l}$ (as well as $\hat {\cal G}_{\bar l}$) corresponds to the structured switched system associated with ${\cal G}_{\rm cact}$ (i.e., preserving the edges in ${\cal G}_{\rm cact}$ and removing those not in ${\cal G}_{\rm cact}$).  Let $S_{i}\subseteq \hat U$ and $T_i\subseteq \hat X_{0}$ be respectively the set of tails and heads of paths in $L_i\doteq \{\hat p^{q_i}_{u_{i0}x_{ik}}: k=1,...,d_i\}$, and $S'_{i}\subseteq \hat U$ and $T'_i\subseteq \hat X_{0}$ be respectively the set of tails and heads of paths in $L'_i\doteq \{\hat p_{u_{i0}'x_{ik}'}: k=1,...,s_i\}$. Then, $w(L)=\prod_{i=1}^dw(L_i)\prod_{i=1}^sw(L_i')$.  By Proposition \ref{property-stem-bud} and the fact that there is a unique path from $u_{i0}'$ to each state vertex of ${\cal G}_{\rm stem}^i$ per $i\in [s]$, as well as the construction of the MDG $\hat {\cal G}_{\bar l}$, we know that there is only one nonzero entry per row and column in $W_{\bar l}(T_i',S_i')$, leading to $\det W_{\bar l}(T_i',S_i')=\pm w(L_i')$. Therefore, $w(L_i')$ cannot be canceled out in $\det W_{\bar l}(T_i',S_i')$, per $i\in [s]$.

Next, we shall show by contradiction that $w(L_i)$ cannot be canceled out in $\det W_{\bar l}(T_i,S_i)$, per $i\in [d]$.
Assume that there is a linking $\tilde L_i$ from $S_i$ to $T_i$ such that $w(\tilde L_i)=w(L_i)$, for some $i\in [d]$. Suppose that the $d_i$ MDG-paths constituting $\tilde L_i$ correspond to $d_i$ input-state walks in ${\cal G}_c$, given by $\tilde p_{u_{i0}x_{i1}},...,\tilde p_{u_{i0}x_{id_i}}$ ($\tilde p_{u_{i0}x_{ij}}$ denotes a walk from $u_{i0}$ to $x_{ij}$). Notice that $w(\tilde L_i)=w(L_i)$ requires that the number of each individual edge contained in $\tilde p_{u_{i0}x_{i1}},...,\tilde p_{u_{i0}x_{id_i}}$ is the same as that in $p_{u_{i0}x_{i1}}^{q_i},...,p_{u_{i0}x_{id_i}}^{q_i}$; moreover, there are one-one correspondences between the color indices of $\tilde p_{u_{i0}x_{i1}},...,\tilde p_{u_{i0}x_{id_i}}$ and of $p_{u_{i0}x_{i1}}^{q_i},...,p_{u_{i0}x_{id_i}}^{q_i}$. Suppose $x_{ii^*}$ is the first vertex in ${\cal G}_{\rm bud}^i$ when walking along the path $p_{u_{i0}\to x_{i1}}$ (recall that $x_{i1}$ is the first vertex in the cycle of ${\cal G}_{\rm bud}^i$ when walking along the path $p_{u_{i0}\to x_{i1}}$), and let $p_{u_{i0}\to x_{ii^*}}$ and $p_{x_{ii^*}\to x_{i1}}$ be subgraphs of $p_{u_{i0}\to x_{i1}}$ such that $p_{u_{i0}\to x_{i1}}=p_{u_{i0}\to x_{ii^*}}\vee  p_{x_{ii^*}\to x_{i1}}$.
Since $p_{u_{i0}\to x_{ii^*}}$ is the unique simple path from $u_{i0}$ to $x_{ii^*}$ in ${\cal G}_{\rm cact}$, $p_{u_{i0}\to x_{ii^*}}$ must be a subgraph of $\tilde p_{u_{i0}x_{ik}}$, per $k\in [d_i]$.
Assume further, for the sake of contradiction, that there is one walk, denoted by $\tilde p_{u_{i0}x_{ik'}}$, $k'\in [d_i]$, that does not reach the cycle $p_{x_{i1}\to x_{i1}}$ (we say $\tilde p_{u_{i0}x_{ik'}}$ reaches $p_{x_{i1}\to x_{i1}}$ if the former contains at least one vertex of the latter). Observe that the total number of edges contained in $p_{u_{i0}x_{i1}}^{q_i},...,p_{u_{i0}x_{id_i}}^{q_i}$ but not belonging to $p_{x_{i1}\to x_{i1}}$ is $d_i|p_{u_{i0}\to x_{i1}}|+\sum_{k=1}^{d_i} |p_{x_{i1}x_{ik}}|< d_i|p_{u_{i0}\to x_{ii^*}}|+2d_i^2$, where $p_{u_{i0}\to x_{i1}}=p_{u_{i0}\to x_{ii^*}}\vee  p_{x_{ii^*}\to x_{i1}}$ and $|p_{x_{ii^*}\to x_{i1}}|,|p_{x_{i1}x_{ik}}|< d_i$ have been used. This indicates that the length of $\tilde p_{u_{i0}x_{ik'}}$ is less than $2d_i^2+|p_{u_{i0}\to x_{ii^*}}|$. However, notice that the length of $p^{q_i}_{u_{i0}x_{ik}}$ per $k\in [d_i]$ is at least $|p_{u_{i0}\to x_{ii^*}}|+q_i\ge 2d_i^2+|p_{u_{i0}\to x_{ii^*}}|$ from (\ref{construct}), so is of $\tilde p_{u_{i0}x_{ik}}$ per $k\in [d_i]$. This contradiction implies that every walk $\tilde p_{u_{i0}x_{ik}}$ per $k\in [d_i]$ must reach the cycle $p_{x_{i1}\to x_{i1}}$, and therefore contains $p_{u_{i0}\to x_{i1}}$ as a subgraph (recall that $p_{u_{i0}\to x_{i1}}$ is the unique simple path from $u_{i0}$ to $x_{i1}$ in ${\cal G}_{\rm cact}$). Notice that the remaining edges of $\tilde p_{u_{i0}x_{ik}}|_{k=1}^{d_i}$, except for those in $p_{u_{i0}\to x_{i1}}$, belong to ${\cal G}_{\rm bud}^i$. Due to the S-disjointness of those edges, given a sequence of color indices, there is at most one walk in ${\cal G}_{\rm cact}$ starting from $x_{i1}$ that corresponds to it. As a result, the only linking from $S_i$ to $T_i$ such that $w(\tilde L_i)=w(L_i)$ is itself, i.e., $\tilde L_i=L_i$. In other words,  $w(L_i)$ cannot be canceled out in $\det W_{\bar l}(T_i,S_i)$, per $i\in [d]$.

%if there is a path from a vertex $u\in S_i$ (resp. $u\in S'_i$) to a vertex $x'\in T_i$ ($x'\in T'_i$), then this path is the unique path from $u$ to $x'$ in $\hat {\cal G}_{\bar l}$. Moreover, there is no path from any $u''\in S_i\backslash \{u\}$ (resp. $u''\in S'_i\backslash \{u\}$) to $x'$, and no path from $u$ to any $x''\in T_i\backslash \{x'\}$ (resp. $x''\in T'_i\backslash \{x'\}$).\footnote{This can be justified by contradiction. Assume that there are paths from the same vertex $u\in S_i$ to two different vertices $x'\in T_i$ and $x''\in T_i$ in $\hat {\cal G}_{\bar l}$. Then, these paths have the same length and the same sequence of color indices (as each copy of the state vertex set is connected to a unique copy of the state vertex set in its lower layer), contradicting the S-disjointness condition. Moreover, if there exist paths from $u'\in S_i$ and from $u''\in S_i$ to the same vertex $x\in T_i$, then there is at least one vertex, say $u'\in S_i$, which is the tail of two paths with different heads in $T_i$. This falls into the first case.  Other cases can be justified similarly.} As a result, there is only one nonzero entry per row and column in $W_{\bar l}(T_i',S_i')$, $i=1,...,s$, so is in $W_{\bar l}(T_i,S_i)$, $i=1,...,d$. %This means $w(L)=\prod_{i=1}^d|\det(W_{\bar l}(T_i,S_i))|\prod_{i=1}^s|\det(W_{\bar l}(T_i',S_i'))|$. %$w(L)=\prod_{i=1}^dw(L_i)\prod_{i=1}^sw(L_i')$

%Similarly, observe that there is a unique path from $u_{i0}$ to $x_{i1}$ in ${\cal G}_{\rm cact}$ and given a sequence of color indices,

As $\{p_{u'_{i0}x'_{ik}}: k\in[s_i]\}$ and $\{p_{u'_{j0}x'_{jk}}: k\in[s_j]\}$ have no common vertices when $i\ne j$, we have $W_{\bar l}(T_i',S_j')=0$ for any $i\ne j$.
Therefore, if there is a linking $\bar L$ such that $w(\bar L)=w(L)$, ${\rm head}(\bar L)={\rm head}(L)$ and ${\rm tail}(\bar L)={\rm tail}(L)$, the only possible case is that $\bigcup_{i=1}^d S_i$ can be partitioned into disjoint subsets $\bar S_i|_{i=1}^d$, such that there exists a $\bar S_i-T_i$ linking, denoted as $\bar L_i$, per $i\in [d]$, satisfying $\prod_{i=1}^dw(L_i)=\prod_{i=1}^dw(\bar L_i)$. Suppose $i^*$ is the smallest $i$ satisfying $\bar S_i\ne S_i$. Then, there is at least one $x'\in T_{i^*}$ and one $u\in \bar S_{i^*}$ but $u\notin S_{i^*}$, such that a path from $u$ to $x'$ exists in $\hat {\cal G}_{\bar l}$. Denote such a path by $p_{u\to x'}$. Observe that the length of $p_{u\to x'}$ is larger than the longest path in the linking $L_{i^*}$. As a result, there is at least one edge in ${\cal G}_{\rm bud}^{i^*}$, denoted as $e$, such that the degree of $w(e)$ in the term $w(\bar L_{i^*})$ is larger than that in $w(L_{i^*})$ (the degree of $w(e)$ in $w(\bar L_{i^*})$ equals the number of occurrences of $e$ in all paths of $\bar L_{i^*}$). Notice also that the factor involving $w(e)$ in other term $w(\bar L_j)$, $j\in \{i^*+1,...,d\}$, if any, is the same as it is in $w(L_j)$, since the path from ${\cal G}_{\rm bud}^{i^*}$ to ${\cal G}_{\rm bud}^j$, $i^*<j$, is unique (if any) in ${\cal G}_{\rm cact}$.
This further leads to that $w(L)\ne w(\bar L)$, yielding that ${\rm sign}(L)w(L)$ cannot be canceled out in (\ref{linking-determinant}). Therefore, $W_{\bar l}$ has full row generic rank, indicating the structural controllability of system (\ref{plant-switched}).
\end{proof}

%Below, we provide an example to illustrate the last argument in step 2 of the above proof.
%\begin{example}
%Consider a single system $(A_1,B_1)$ shown in Fig. xx.
%\end{example}
						
						It can be seen that the graph ${\cal G}_{\rm cact}$ constructed in the proof of Theorem \ref{main-theorem}, which consists of the union of vertex-disjoint generalized stems and buds, as well as the unique paths (in ${\cal G}_{\rm cact}$) from an input vertex to each vertex of the cycle per generalized bud, coincides with the {\emph{cactus}} concept introduced in Lin's work \cite{C.T.1974Structural} when $N=1$. Hence, we call ${\cal G}_{\rm cact}$ a generalized cactus. {See Fig. \ref{example-cactus}(a) for an illustration of a generalized cactus.} It follows that ${\cal G}_{\rm cact}$ serves as the fundamental structure of ${\cal G}_c$ that preserves structural controllability of the original system (that is, edges not in ${\cal G}_{\rm cact}$ can be deleted from ${\cal G}_c$ so that the remaining subgraph corresponds to a structurally controllable system).\footnote{We remark that \cite{C.T.1974Structural} imposed some further constraints on the cactus, so that it is the {\emph{minimal structure}} that preserves structural controllability, i.e., removing any edge from it will result in a structurally uncontrollable structure.} %We leave characterizing such minimal structure for structural controllability of switched systems based on generalized cacti as future work.
						The following corollary is immediate from the proof of Theorem \ref{main-theorem}, revealing that the number of state vertices covered by a generalized cactus configuration provides a lower bound for the generic dimension of controllable subspaces.
						\begin{corollary}\label{main-lower-bound}
							Suppose $X_s\subseteq X$ is covered by a generalized cactus configuration. Then, ${\rm grank}{\cal R}\ge |X_s|$.
						\end{corollary}

							\begin{theorem}\label{grank-theorem} The following conditions are equivalent:
								\begin{itemize}
									\item[(a)] Every state vertex $x_i\in X$ is input-reachable in ${\cal G}_c$ (a.i), and ${\rm grank}[A_1,...,A_N,B_1,...,B_N]=n$ (a.ii).
									\item[(b)] ${\cal G}_c$ contains a collection of vertex-disjoint generalized stems and generalized buds that cover $X$ (i.e., ${\cal G}_c$ contains a generalized cactus configuration covering $X$).
									\item[(c)] System (\ref{plant-switched}) is structurally controllable.
								\end{itemize}
							\end{theorem}
							
							\begin{proof} (b)$\Rightarrow$(a): By definition, every vertex in a generalized stem or bud is input reachable, yielding condition (a.i). Moreover, observe that all edges in a generalized stem or bud are S-disjoint. The set of edges of a generalized cactus configuration covering $X$ is S-disjoint with size $n$ immediately. By Lemma \ref{s-disjoint-edge}, condition (a.ii) holds.
								
								%	Let $\{p_1,...,p_n\}$ be a generalized cactus walking with size $n$ in ${\cal G}_c$.  Since every $v\in X$ is the head of an input-state walk, every $v$ should be input-reachable, leading to condition (a.i). Take the last edge $e_i$ of $p_i$ per $i\in \{1,..,n\}$. It is easy to see that $e_1,...,e_n$ are $n$ S-disjoint edges, leading to condition (a.ii) via Lemma \ref{s-disjoint-edge}.
								
								(a)$\Rightarrow$(b): Suppose conditions (a.i) and (a.ii) hold. By Lemma \ref{s-disjoint-edge}, there exist $n$ S-disjoint edges $E_S$ in ${\cal G}_c$. We are to show that a collection of vertex-disjoint generalized stems and buds can be constructed from $E_S$. To begin with, a generalized stem ${\cal G}_{\rm stem}=(V_{\rm stem},E_{\rm stem})$ is constructed as follows. Let $e_1\in E_S$ be such that ${\rm tail}(e_1)\in U$ (if such an edge does not exist, then no generalized stem can be found). Add $e_1$ to $E_{\rm stem}$. Then, find {\emph{all}} edges of $E_2\doteq \{e\in E_S: {\rm tail}(e)={\rm head}(e_1)\}$, and add them to $E_{\rm stem}$. Next, add all edges of $E_3\doteq \{e\in E_S: {\rm tail}(e)={\rm head}(e_2), e_2\in E_2\}$ to $E_{\rm stem}$. Repeat this procedure until the set $E_{i+1}=\{e\in E_S: {\rm tail}(e)={\rm head}(e_i), e_i\in E_i\}$ is empty for some $i$ (let $E_1=\{e_1\}$). Let $V_{\rm stem}$ be the set of end vertices of edges in $E_{\rm stem}$. Then, it can be verified that all three conditions in Definition \ref{def-general-stem} are satisfied for ${\cal G}_{\rm stem}=(V_{\rm stem}, E_{\rm stem})$. Hence, ${\cal G}_{\rm stem}$ is a generalized stem. Other generalized stems can be constructed from $E_S\backslash E_{\rm stem}$ similarly.
								
								After finding all generalized stems from $E_{S}$, let $E^0_{S}$ be the subset of $E_S$ by removing all edges belonging to the generalized stems. A generalized bud ${\cal G}_{\rm bud}=(V_{\rm bud},E_{\rm bud})$ can be constructed as follows. Pick an arbitrary $e_1\in E^0_{S}$ and add it to $E_{\rm bud}$. Let $E_1=\{e_1\}$ and $E^1_S=E^0_S\backslash E_1$. Find $E_2=\{e\in E^1_S: {\rm head}(e)={\rm tail}(e_1), {\rm or}\ {\rm tail}(e)={\rm head}(e_1)\}$, and add $E_2$ to $E_{\rm bud}$. Let $E_S^2=E_S^1\backslash E_2$.  Next, find $E_3=\{e\in E^2_S: {\rm head}(e)={\rm tail}(e_2), {\rm or}\ {\rm tail}(e)={\rm head}(e_2), e_2\in E_2\}$, add $E_3$ to $E_{\rm bud}$, and update  $E_S^3=E_S^2\backslash E_3$. Repeat this procedure until the set $E_{i+1}=\{e\in E^i_S: {\rm head}(e)={\rm tail}(e_i), {\rm or}\ {\rm tail}(e)={\rm head}(e_i), e_i\in E_i\}$ is empty. Let $V_{\rm bud}$ be the set of end vertices of edges in $E_{\rm bud}$. It turns out that there is at least one cycle in ${\cal G}_{\rm bud}=(V_{\rm bud},E_{\rm bud})$. If not, since every vertex $x\in V_{\rm bud}$ has exactly one ingoing edge in $E_{\rm bud}$, the above-mentioned procedure cannot terminate. On the other hand, if there are two or more cycles in ${\cal G}_{\rm bud}$, then at least one vertex has two or more ingoing edges, a contraction to the S-disjointness condition. Therefore, ${\cal G}_{\rm bud}$ is a generalized bud. Other generalized buds can be found similarly.
								
								(c)$\Rightarrow$(a): From Proposition \ref{determinant}, if system (\ref{plant-switched}) is structurally controllable, i.e., ${\rm grank}W_{n}=n$, then there is a $\hat U-\hat X_{0}$ linking of size $n$ in ${\hat {\cal G}}_{n}$. By the construction of ${\hat {\cal G}}_{n}$, every state vertex $x_{i0}^{00}\in \hat X_{0}$ is the head of a path from $\hat U$, which leads to the necessity of condition (a.i). Moreover, there are $n$ vertex-disjoint edges between $\hat U_{0}\cup \hat X_{1}$ and $\hat X_{0}$, which implies the necessity of condition (a.ii) via Lemma~\ref{s-disjoint-edge}.
								
								As (b)$\Rightarrow$(c) has been proved in Theorem \ref{main-theorem}, the equivalence among (a), (b) and (c) is immediate.
							\end{proof}
							
							\begin{remark}% \subsection{Examples}
								The proof `(a)$\Leftrightarrow$(b)' above implies that a generalized cactus configuration covering $X$ can be uniquely determined by a set of $n$ S-disjoint edges, and vice versa (given that all $x_i\in X$ are input-reachable). {Note that the input-reachability condition (condition a.i) in ${\cal G}_c$ can be verified using a standard depth-first search algorithm, with a complexity of $O(|X\cup U|+|E_{XX}\cup E_{UX}|)$. To test whether ${\rm grank}[A_1,...,A_N,B_1,...,B_N]=n$ (i.e., condition a.ii), one can employ maximum matching algorithms to find a maximum cardinality matching in the bipartite graph associated with the structured matrix $[A_1,...,A_N,B_1,...,B_N]$, incurring $O(\sqrt{nN}M)$ complexity, where $M$ denotes the number of nonzero entries in $[A_1,...,A_N,B_1,...,B_N]$; see \cite[Sec 3.3]{LiuStructural} for details. Thus, the verification of the structural controllability conditions outlined in Theorem \ref{grank-theorem} can be achieved within polynomial time.}
							\end{remark}

							\begin{example}
								Consider a switched system whose ${\cal G}_c$ is given in Fig. \ref{example-cactus}(a). It turns out that ${\cal G}_c$ can be spanned by the union of a generalized stem Fig. \ref{example-cactus}(b) and a generalized bud Fig. \ref{example-cactus}(c). From Theorem \ref{grank-theorem}, this switched system is structurally controllable.
							\end{example}
							
	\begin{example} \label{up-low-rank-example}
	Consider a switched system with $N=2$ and $n=10$. The colored union graph ${\cal G}_c$ is given in Fig. \ref{example-subspace}(a), where there is no multiple edge (thus the subsystem digraphs ${\cal G}_1$ and ${\cal G}_2$ can be uniquely extracted from it). Let $(A_1,B_1)$ and $(A_2, B_2)$ be respectively the system matrices corresponding to ${\cal G}_1$ (solid edges) and ${\cal G}_{2}$ (dotted edges). It can be verified that ${\rm grank}[A_1,A_2,B_1,B_2]=9$, implying that this system is not structurally controllable via Theorem \ref{grank-theorem}. Furthermore, Fig. \ref{example-subspace}(b) shows that vertices $\{x_1,...,x_6,x_9,x_{10}\}$ can be covered by a generalized cactus configuration. It follows from Corollary \ref{main-lower-bound} that ${\rm grank}{\cal R}\ge 8$. A lower bound of ${\rm grank}{\cal R}\ge 6$ can also be obtained from Fig. \ref{example-subspace}(c), which shows that ${\cal G}_c$ contains a {\emph{conventional}} cactus configuration covering $\{x_1,...,x_5,x_9\}$. A further inspection on the MDG $\hat {\cal G}_{\bar l}$ yields that any $\hat U_{}-\hat X_{0}$ linking in $\hat {\cal G}_{\bar l}$ for any $\bar l\ge 2$ has a size at most $8$. Combining it with the lower bound from Fig. \ref{example-subspace}(b), we obtain ${\rm grank}{\cal R}= 8$.
	\begin{figure}
		\centering
		\includegraphics[width=1.8in]{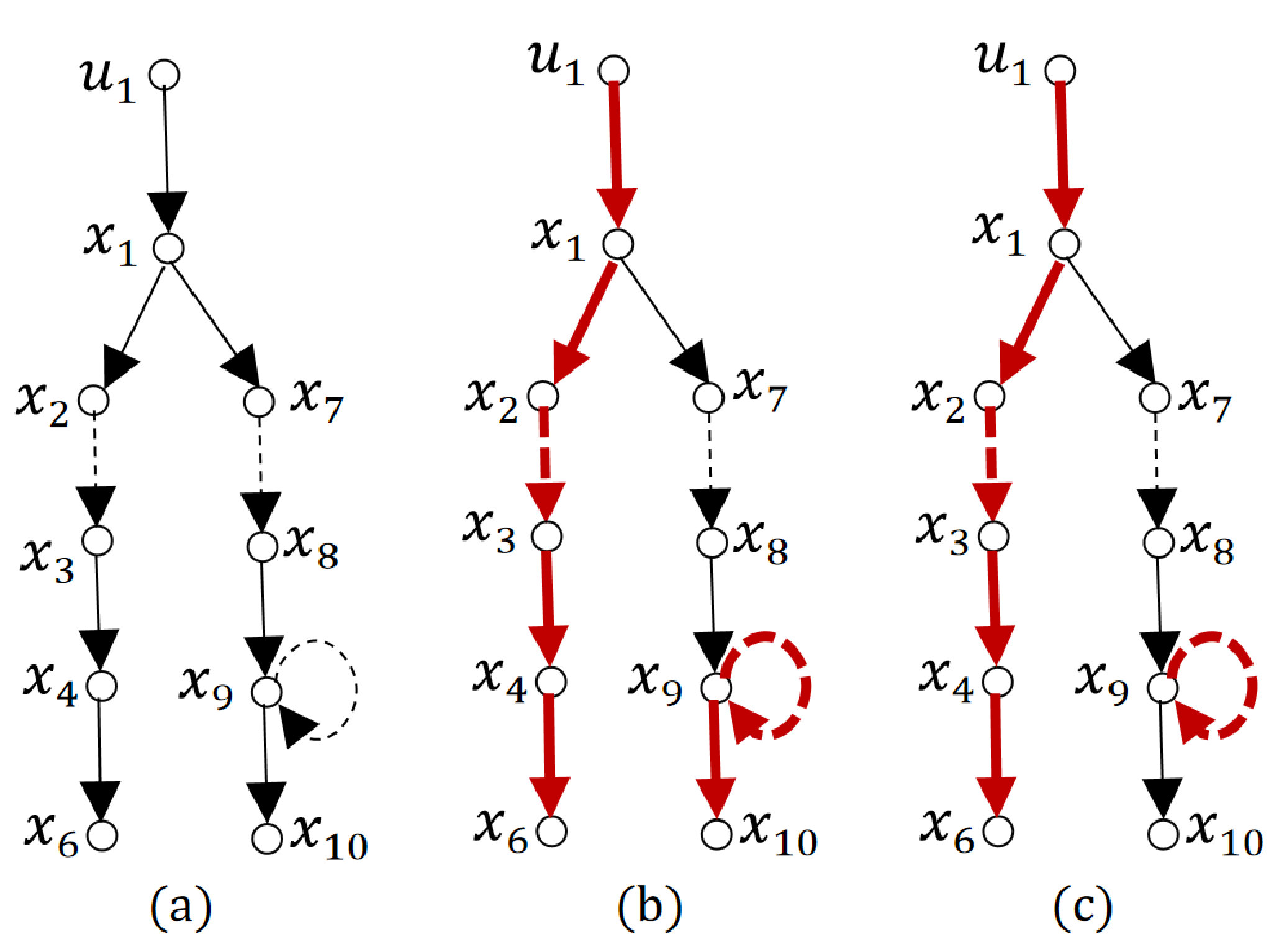}\\
		\caption{(a): ${\cal G}_c$ (no multiple edges) of a switched system with $N=2$ and $n=10$. (b): a generalized cactus configuration in ${\cal G}_c$ (depicted in bold red lines). (c): a conventional cactus configuration in ${\cal G}_c$ (bold red lines).}\label{example-subspace}
	\end{figure}
\end{example}

Example \ref{up-low-rank-example} shows that the lower bound for ${\rm grank}{\cal R}$ provided by Corollary \ref{main-lower-bound} can be tighter than the one provided by the conventional cactus configuration covering. In addition, the upper bound provided by Proposition \ref{determinant} can be tighter than the one provided by ${\rm grank}[A_1,...,A_N,B_1,...,B_N]$.
Notice that when ${\rm grank}{\cal R}=n$, or when $N=1$ (i.e., for LTI systems), these two bounds coincide, both equaling the exact generic dimension of controllable subspaces \cite{poljak1990generic}. {How to characterize ${\rm grank}{\cal R}$  when $N\ge 2$ is yet open.}
							\section{Conclusions}

							In this paper, we have investigated the structural controllability of switched continuous-time systems. By introducing new graph-theoretic concepts such as multi-layer dynamic graph, generalized stem, bud, and cactus, we provide a novel generalized cactus based criterion for structural controllability, which extends Lin's graph-theoretic condition to switched systems for the first time. This not only fixes a gap in the existing literature regarding the proof of a pivotal criterion for structural controllability, but also yields a lower bound for the generic dimension of controllable subspaces. Our future work will consist of characterizing the generic dimension of controllable subspaces based on the proposed generalized cacti and extending these results to switched discrete-time systems. %{Notice that unlike switched continuous-time systems, verifying controllability of switched discrete-time systems usually needs to check an infinite number of switching paths \citep[Theo 1, 2]{ge2001reachability}, which brings additional challenge for the structural controllability problem.}
							% and reversible switched discrete time systems %characterizing the minimal structure that can preserve structural controllability based on the proposed generalized cacti,
							\section*{Acknowledgment}
							%	Suggestions from Prof. B. M. Chen of the Chinese University of HongKong on this topic are highly appreciated.
							Feedback from Prof. B. M. Chen of the Chinese University of HongKong on this manuscript is highly appreciated.

					}}
					
					%	{\bf Claim:} The following three numbers are equivalent:
					%	
					%	(1) The generic dimension of reachable (resp. controllable) subspace of the switched system (\ref{plant-switched});
					%	
					%	(2) The maximum size of $(U,V_{end})$ linkings in the dynamic graph ${\cal G}_{s}$;
					%	
					%	(3) The maximum number of state vertices that are covered by a generalized cacti configuration with distinct length sequences.
					
					%	\section*{\refname}
					\bibliographystyle{elsarticle-num}
					{\footnotesize
						\bibliography{yuanz3}
					}
				\end{document}